\documentclass[usenatbib]{mn2e}
\input psfig.sty
\usepackage{epsfig}
\usepackage{amsmath,amssymb}
\usepackage{psfig}

\renewcommand{\vec}[1]{{\mathbf #1}}

\title[Simulating fluids using SPH and grid techniques]
      {Fundamental differences between SPH and grid methods}
      \author[Oscar Agertz et al.]
{\parbox[t]{\textwidth}{Oscar Agertz$^1$\thanks{agertz@physik.unizh.ch}, 
Ben Moore$^1$,
Joachim Stadel$^1$, 
Doug Potter$^1$, 
Francesco Miniati$^2$,
Justin Read$^1$,
Lucio Mayer$^2$, Artur Gawryszczak$^3$, Andrey Kravtsov$^4$, Joe Monaghan$^5$, \AA ke Nordlund$^6$, Frazer Pearce$^7$, Vincent Quilis$^8$, Douglas Rudd$^4$, Volker Springel$^{9}$, James Stone$^{10}$, Elizabeth Tasker$^{11}$, Romain Teyssier$^{12}$, James Wadsley$^{13}$, Rolf Walder$^{14}$}\vspace*{3pt}\\
$^1$ Institute for Theoretical Physics, University of Z\"urich, CH-8057 Z\"urich, Switzerland \\
$^2$ Department of Physics, Institute f\"ur Astronomie, ETH Z\"urich, CH-8093 Z\"urich, Switzerland\\
$^3$ Nicolaus Copernicus Astronomical Centre, Bartycka 18, Warsaw, PL-00-716, Poland\\  
$^4$ Department of Astronomy \& Astrophysics, The University of Chicago, Chicago, IL 60637 USA\\
$^5$ School of Mathematical Sciences, Monash University, Clayton 3800, Australia\\
$^6$ Niels Bohr Institute, Copenhagen University, Juliane Maries Vej 30, DK-2100 K\o benhavn \O, Denmark\\  
$^7$ School of Physics and Astronomy, University of Nottingham, University Park, Nottingham NG7 2RD, UK\\
$^8$ Departemento de Astronomia y Astrofisica, Universidad de Valencia, 46100, Burjasott, Valencia, Spain\\
$^{9}$ Max-Planck-Institute for Astrophysics, Karl-Schwarzschild-Str. 1, 85740 Garching, Germany\\
$^{10}$ Department of Astrophysical Sciences, Princeton University, Princeton, NJ 08544\\
$^{11}$ Department of Astronomy, Columbia University, New York, NY 10027\\
$^{12}$ Service d'Astrophysique, CEA/DSM/DAPNIA/SAp, Centre d'Etudes de Saclay, L'orme des Merisiers, 91191 Gif-sur-Yvette Cedex, France\\
$^{13}$ Department of Physics and Astronomy, McMaster University, Hamilton, Ontario, L88 4M1, Canada\\
$^{14}$ Institute f\"ur Astronomie, ETH Z\"urich, CH-8092 Z\"urich, Switzerland\\
 }
\date{\today} 

\begin{document}

\maketitle

\begin{abstract}
We have carried out a hydrodynamical code comparison study of interacting multiphase fluids.
The two commonly used techniques of grid and smoothed particle hydrodynamics (SPH) show striking differences in their ability to model processes that are fundamentally important across many areas of astrophysics. 
Whilst Eulerian grid based methods are able to resolve and treat important dynamical instabilities, such as Kelvin-Helmholtz or Rayleigh-Taylor, these processes are poorly or not at all resolved by existing SPH techniques. We show that the reason for this is that SPH, at least in its standard implementation, introduces spurious pressure forces on particles in regions where there are steep density gradients. This results in a boundary gap of the size of the SPH smoothing kernel over which information is not transferred. 
\end{abstract}


\section{Introduction}

The ability to numerically model interacting fluids is essential to many areas of astrophysics and other disciplines. From the formation of a star and its proto-planetary disk to galaxies moving through the intra-cluster medium, dynamical instabilities such as Kelvin-Helmholz (KH) and Rayleigh-Taylor (RT) play a fundamental role in astrophysical structure formation. 
Most popular hydrodynamical methods can be divided into two classes: techniques following the gas using Eulerian grids \citep[e.g.][]{laney_98,leveque98} and those which follow the Lagrangian motions of gas particles such as `Smoothed Particle Hydrodynamics' (SPH) \citep{monaghan92}.  Grid based techniques solve the fluid dynamical equations by calculating the flux of information through adjacent cells, SPH techniques calculate the gas properties on each particle by averaging over its nearest neighbours .
Due to the extensive use of these techniques, it is interesting to carry out code comparison studies on well defined problems that test their ability to follow the basic gas physics they are designed to simulate.

Our test problem is to follow a dense cold gas cloud moving through a low density hot medium. This is specifically designed to capture the same physical processes that occur during the formation and evolution of astrophysical structures. We will also study the shearing motion of two fluids of different densities to elucidate the problems that we find with this test. 
Similar configurations, including shockwave interaction with clouds, have been studied by e.g. \cite{murray93}, \cite{klein94}, \cite{vietri97}, \cite{mori00}. However we are not aware of a direct comparison between simulation methods in this context. Differences were found in the literature between different studies of the same problem. For example, SPH studies of galaxy-cluster interactions by \cite{abadi99} found that only half the inter-stellar medium was removed from the galaxy. Using a grid based calculation of the same initial conditions, \cite{quilis00} found that all the gas could be removed and attributed this to the high resolution shock capturing ability of their Eulerian code.

\section{The blob test}	

\label{sec:prob}
\begin{figure}
\psfig{file=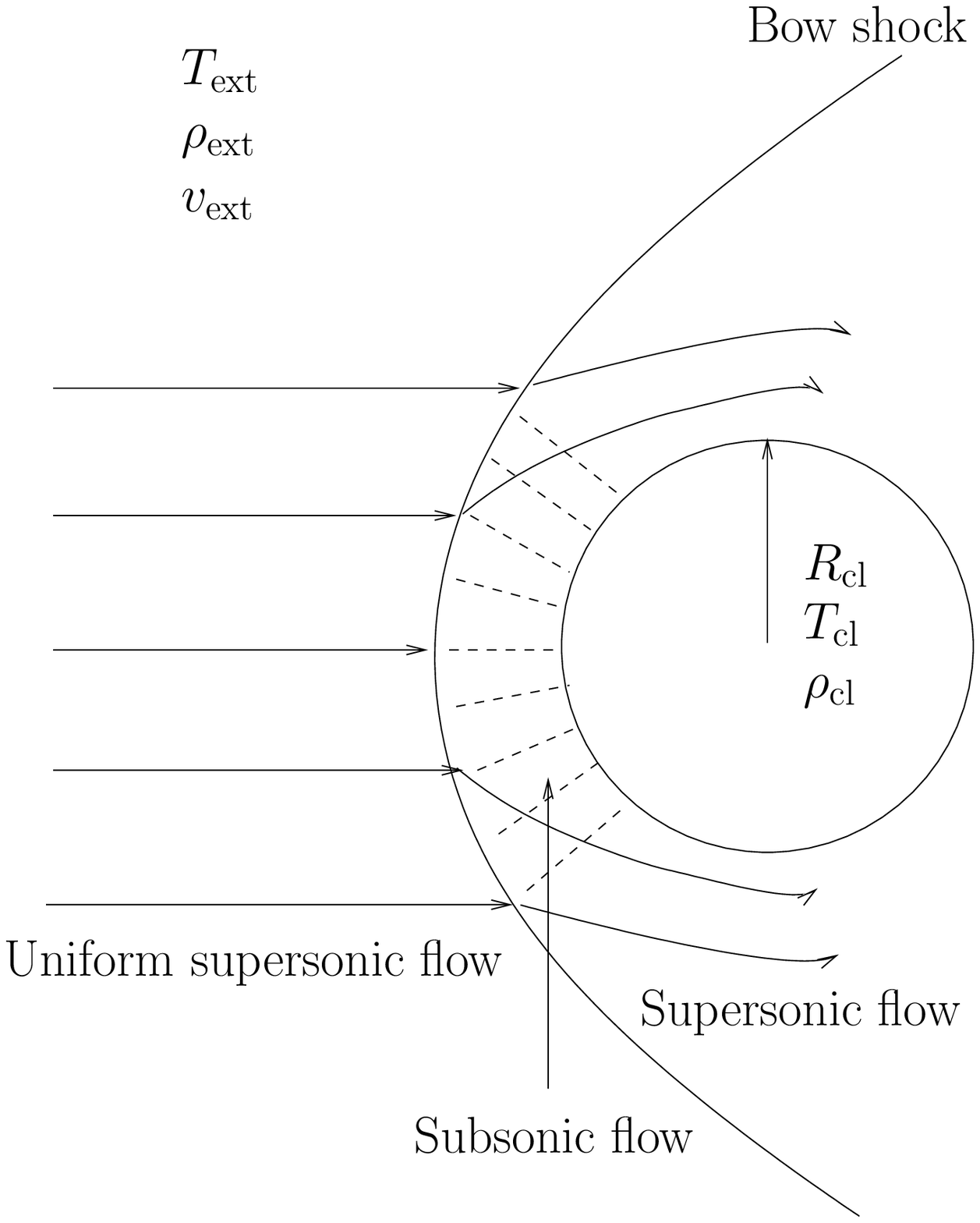,height=220pt,bbllx=70bp,bblly=230bp,bburx=545bp,bbury=762bp,clip=}
\caption[]{Illustration of the blob test. The external medium, which initially is in pressure equilibrium with the cloud, travels with a supersonic velocity creating a bow shock in front of the cloud. The post shock flow is subsonic until the smooth flow accelerates and again obtains supersonic speed on the lateral sides of the cloud.} 
\label{fig:linear}
\end{figure}

A schematic view of the blob test problem can be seen in Fig. \,\ref{fig:linear}. A spherical cloud of gas is placed in a wind tunnel with periodic boundary conditions. The ambient medium is ten times hotter and ten times less dense than the cloud so that it is in pressure equilibrium with the latter. We will refer to this initial density contrast between the cloud and the medium as $\chi_{\rm{ini}}$. All of the gas is atomic hydrogen with molecular weight $m_{\rm{mol}}=1.0$ and an adiabatic index $\gamma=5/3$.

This setup will investigate how different simulation codes handle typical astrophysical processes important for multi-density and multi-phase systems, such as ram-pressure stripping and fragmentation through KH and RT instabilities.

\begin{figure}
\psfig{file=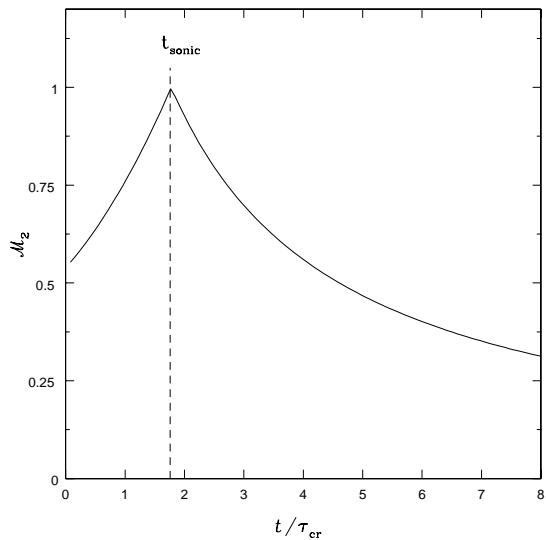,height=220pt}
\caption[]{We plot the Mach number of the flow directly downstream of the shock on the symmetry axis of the cloud. The flow speed increases due to the weakened shock strength up to $t_{\rm{sonic}}$ where the relative motion of the cloud and wind turns subsonic.}
\label{fig:mach}
\end{figure}
\section{Analytical expectations}
Although the nonlinear stages of the KH and RT instabilities
cannot be fully described analytically, we can still use analytic
arguments to estimate the characteristic disruption timescale
for the cloud.

In order to specify our problem we characterize the external medium
with a sound speed $c_{\rm s}$ and assign it an initial velocity
$v=\mathcal{M} c_{\rm s}$ with Mach number $\mathcal{M}=2.7$. Furthermore, we place the cloud
initially at rest in the computational domain. Since the wind is supersonic, a bow shock will form in front of the cloud with the post shock properties given by the Rankine-Hugonoit shock jump conditions. Because the cloud is accelerated by the wind, we will from now on perform all of our calculations \emph{in the rest frame of the bow shock}, referring to pre-shock quantities with the subscript 1 and post-shock with 2. The shock conditions for the density, velocity and Mach number are \citep[e.g.][]{shu92}
\begin{equation}
\label{machjump}
\frac{\rho_2}{\rho_1}=\frac{v_1}{v_2}=\frac{(\gamma+1)\mathcal{M}_1^2}{(\gamma+1)+(\gamma-1)(\mathcal{M}_1^2-1)}
\end{equation}
\begin{equation}
\label{machjump2}
\mathcal{M}_2^2 = \frac{2+(\gamma-1)\mathcal{M}_1^2}{2\gamma \mathcal{M}_1^2-(\gamma-1)}
\end{equation}
Formally we would take the obliqueness of the bow shock into account but for simplicity we will only consider the flow that enters at the symmetry axis of the cloud. 

The cloud acceleration can be approximated by considering the maximum area that can gain momentum from the ambient flow. This implies that all gas in a cylinder in front of the cloud transfers momentum leading to an acceleration 
\begin{equation}
\label{eq:acceleration}
a_{\rm{cl}} \sim \dot{v}_1 \sim \frac{\rho_{\rm{ext}}\pi R_{\rm{cl}}^2v_{1}^2}{M_{\rm{cl}}}.
\end{equation}
Integrating this equation leads us to the evolution of the pre-shock velocity
\begin{equation}
\label{eq:velocity} 
v_1(t)=\frac{l}{(t+l/v_{\rm{ext}})},
\end{equation}
where $l$ is a characteristic length given by $l=M_{\rm{cl}}/2\pi R_{cl}^2\rho_{\rm{ext}}$.  By using Eq. \,\ref{eq:velocity} to calculate the pre-shock Mach number together with Eq. \,\ref{machjump2} we can obtain a qualitative understanding of the post-shock velocity. This velocity is crucial for the stability of the cloud surface and, as we will show in section \,\ref{sect:KH}, for the destruction of the cloud itself. The evolution of the post-shock Mach number $\mathcal{M}_2$ is given by
\begin{equation}
\mathcal{M}_2^2= \left\{\begin{array}{ll}
\frac{2+(\gamma-1)(v_1/c_{\rm{s}})^2}{2\gamma(v_1/c_{\rm{s}})^2-(\gamma-1)} & \textrm{for $t<t_{\rm sonic}$} \\
(v_1/c_{\rm{s}})^2 & \textrm{for $t>t_{\rm sonic}$} \\
\end{array}\right.
\label{fig:postmach}
\end{equation}
Here $t_{\rm{sonic}}$ is the time at which $\mathcal{M}_1=\mathcal{M}_2=1$ and the shock disapperars. After this point, gas freely streams towards the cloud and the Mach number decreases only due to the continued acceleration. Notice that for $t<t_{\rm{sonic}}$, $\mathcal{M}_2 < 1$, even for $\mathcal{M}_1 = v_1/c_s \rightarrow \infty$. This means that behind the shock, the flow will always be subsonic and we expect instabilities to grow there. For $t \rightarrow \infty$, $\mathcal{M}_2 \rightarrow 0$ and the cloud will eventually be co-moving with the background flow. The evolution of the postshock Mach number is shown in Fig.\,\ref{fig:mach} in terms of the so-called ``crushing time'' defined as, in our notation,
\begin{equation}
\tau_{\rm cr}=\frac{2R_{\rm{cl}}\chi^{1/2}}{v_{\rm{1}}},
\label{fig:crushtime}
\end{equation}
where $\chi$ is the density contrast between the cloud and the external medium. This is a natural timescale supersonic cloud evolution. We will naively use $\chi=\chi_{\rm{ini}}=10$ and $v_1=v_{\rm ext}$, representing our initial condition. During the interval of $\tau_{\rm{cr}}$ a bowshock is formed and the shocked gas will form a smooth flow around the cloud, reaching supersonic speed at the points indicated in Fig. \,\ref{fig:linear}. Beyond this region we expect to see a turbulent boundary layer forming which transports material off the surface. The cloud will compress along the line of motion due to an internal shock wave generated by the external gas. From Bernoulli's theorem we know that the pressure is low on the lateral sides which causes an overspilling of the cloud due to the high inner pressure of the compressed cloud. This causes mass loss irrespective of any instability.

\subsection{The Kelvin-Helmholtz instability}
\label{sect:KH}
\begin{figure}
\psfig{file=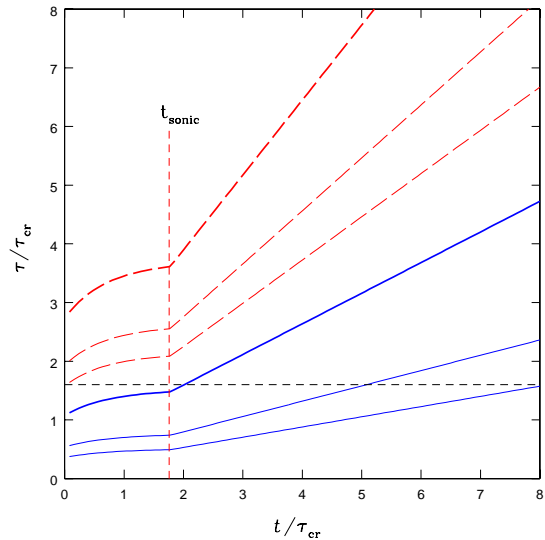,height=220pt}
\caption[]{The time dependence of the growth rates of KH (solid, blue lines) and RT (dashed, red lines) instabilities. The lines represent different sizes of perturbation wavelengths: $R_{\rm{cl}}$ (thick), $R_{\rm{cl}}/2$ (middle) and $R_{\rm{cl}}/3$ (thin).} 
\label{fig:KHgrowth}
\end{figure}

Kelvin-Helmoltz instabilities (KHI) occur when velocity shear is present at the interface between two fluids.
The importance of the KHI, in the context of gas cloud stability, has been studied by many authors e.g. \cite{nulsen82}, \cite{murray93}, \cite{vietri97}, \cite{mori00}. 

Neglecting gravity, the dispersion relation of the KHI, in the notation of our setup, for an incompressible fluid is \citep{chandrasekhar61}
\begin{equation}
w=k\frac{(\rho_{\rm{2}}\rho_{\rm{cl}})^{1/2}v_2}{(\rho_{\rm{2}}+\rho_{\rm{cl}})}\approx\frac{k v_2}{\chi^{1/2}},\end{equation}
where $k$ is the wavenumber of the instability and the last approximation holds for $\chi\gg1$. The characteristic growth time for the KHI is then
\begin{equation}
\label{eq:KHI}
\tau_{KH}\equiv\frac{2\pi}{w}=\frac{(\rho_{\rm{2}}+\rho_{\rm{cl}})}{k(\rho_{\rm{2}}\rho_{\rm{cl}})^{1/2}v_2}\approx\frac{2\pi\chi^{1/2}}{k v_2}.
\end{equation}
By naively using the post-shock quantities of Eq.\,\ref{machjump} and our choice of cloud parameters, we can calculate an approximate time dependence of the KH instability, which is shown in Fig. \,\ref{fig:KHgrowth}, (blue, solid lines) for perturbations of size $R_{\rm{cl}}$ (thick), $R_{\rm{cl}}/2$ (middle) and $R_{\rm{cl}}/3$ (thin). Small scale instabilities grow faster due to the $\tau_{\rm{KH}}\sim k^{-1}$ relation. The first modes to grow are the shortest. Their growth will act to widen the interface between the shearing layers, hence dampening the growth of modes smaller than the thickness of the interface \citep{chandrasekhar61}. The fastest growing modes are now those that are equal to the thickness of the interface. As this process continues, the mode responsible for the cloud destruction is that which is comparable to the size of the cloud itself: $k_{\rm{cl}} \sim 2\pi/R_{\rm{cl}}$ \citep{nulsen82,murray93}. 

The instability growth time is always larger than the cloud crushing time. The horizontal line at $\tau=1.6\,\tau_{\rm cr}$ in Fig. \,\ref{fig:KHgrowth} indicates roughly the time at which the $k_{\rm{cl}}$ KH mode should have grown fully.  We will from now on refer to this time as $\tau_{\rm{KH}}$.

Note that cloud compressibility can be taken into account when calculating the KH growth time \citep[see][]{vikhlinin01}, but was omitted for simplicity. Also note that in certain more physically motivated situations with external gravitational fields, self gravity, physical viscosity, magnetic fields, radiation etc., the KHI is modified and is damped in many cases \citep[e.g.][]{murray93,vietri97,miniati99,gregori00}. 

\subsection{The Rayleigh-Taylor instability}

Rayleigh-Taylor instabilities (RTI) occur when a denser fluid is accelerated by a less dense fluid, or when a heavier fluid displaces a lighter fluid. The cloud is accelerated with respect to the background and we expect RT instabilities to develop. 
The dispersion relation for the RTI is \citep{chandrasekhar61}

\begin{equation}
|w^2|=k^\prime a\left(\frac{\rho_{\rm{cl}}-\rho_{\rm{ext}}}{\rho_{\rm{cl}}+\rho_{\rm{ext}}}\right)\approx k^\prime a,
\end{equation}
where the last approximation is valid for $\chi\gg 1$. The KHI, which results from shearing flows, has a 2D geometry, and can be descibed by as single wavevector $k$. By constrast, the RTI necessarily has a 3D geometry and must be described by a {\it vector} wavelength, $\underline{k}'= (k_1,k_2)$, of magnitude: $k^\prime=\sqrt{k_1^2+k_2^2}$.  The acceleration on the surface can be assumed to be $a=\epsilon a_{\rm{cl}}$,  where  $a_{\rm{cl}}$ is given by Eq.\,\ref{eq:acceleration} and $\epsilon$ is an efficiency factor. Note that it is very difficult to analytically determine the efficiency of the momentum transfer from the external medium onto the cloud. By using $\epsilon=1$ we will get a lower limit on $\tau_{\rm{RT}}$.

Fig. \,\ref{fig:KHgrowth} shows, for our choice of parameters, the characteristic growth times for RT instabilities (red, dashed lines) of size $R_{\rm{cl}}$ (thick), $R_{\rm{cl}}/2$ (middle) and $R_{\rm{cl}}/3$ (thin), demonstrating that $\tau_{\rm{KH}}<\tau_{\rm{RT}}$ for large instabilities. The largest mode grows very slowly and is probably not important in this type of problem. However, we expect that a fast growing small-scale RT instability should develop on the cloud front, especially on the axis of symmetry as the flow rams into the stagnation point. Complicated mixtures of KHI and RTI during later evolution is also expected until the cloud becomes fully comoving with the flow.

\section{Numerical simulations}
\label{sect:sim}

Our numerical simulations solve the Euler equations which neglect physical viscosity and radiative processes; we assume a perfect gas equation of state. Away from shocks, the evolution is strictly adiabatic. This means the gas can only undergo heating and cooling by adiabatic compression or expansion, or by irreversible heating at shocks. In order to isolate the differences in hydrodynamic solvers, we neglect the self gravity of the gas.

\subsection{Initial conditions}
\label{sect:IC}
The initial conditions (IC) for the blob test are set up in the following way: we use a periodic simulation box of size, in units of the cloud radius $R_{\rm cl}$, $\{L_x,L_y,L_z\}=\{10,10,40\}$ where we center the cloud at $\{x,y,z\}=\{5,5,5\}$. The ICs are generated by randomly placing equal mass particles to obtain the the correct densities and cloud radius. Using an SPH code, the system is evolved and allowed to relax to obtain pressure equilibrium. Velocities are added to particles that are in the surrounding low dense ambient medium. One could smoothly increase the velocities to be more faithful to astrophysical situations, but this more violent start together with particle noise serves as the initial seed for surface instabilities of the cloud. Formally this can be seen as a triggering of small scale RT and Richtmyer-Meshkov instabilities\footnote[1]{The Richtmyer-Meshkov instability occurs when a contact discontinuity gets shocked or rapidly accelerated. This generates vorticity and structures similar to those of RT \citep[e.g.][]{inogamov99}.}.

This particle setup is used as IC for the SPH simulations. The ICs for the grid simulations are obtained by smoothing the gas quantities (density, temperature and velocities), using the standard SPH kernel with 32 nearest neighbours, and then mapping these on to a uniform grid. In this way the noise introduced by using discrete particles in the SPH simulations is also present in the grid IC. As we will argue below, the key parameters to study are those connected with the resolution and strength of artificial viscosity therefore our parameter space studies will focus on the effect of these.

\subsection{The codes}
The simulation was carried out with about a dozen different independent simulation codes. Since all the grid codes gave consistent results, and similar for the SPH codes, we shall just present the detailed analysis of a selection of these codes which are summarized in Table \,\ref{table:simsummary}. 
Here we give a brief description of these codes and the methods used for solving the hydrodynamical equations:

\subsubsection{ART (AMR)} 
ART (Adaptive Refinement Tree) is a $N$-body$+$gasdynamics AMR code \citep[][]{kravtsov99, kravtsov_etal02}.
The ART code uses second-order shock-capturing Godunov-type solver
\citep{colella_glaz85} to compute numerical fluxes of gas variables
through each cell interface, with ``left'' and ``right'' states
estimated using piecewise linear reconstruction \citep{vanleer79}.
This is a monotone method that is known to provide good results for a
variety of flow regimes and resolves shocks within
$\approx 1-2$ cells.  A small amount of dissipation in the form of
artificial diffusion is added to numerical fluxes
\citep{collelawoodward:ppm}, as is customary in the shock-capturing
codes. The details of the flux evaluation and summation on mesh
interfaces can be found in
\citet{khokhlov:ftt}. In the simulations presented in this paper, 
a new distributed MPI version of the ART code developed by 
Douglas Rudd and Andrey Kravtsov was used. 

\subsubsection{CHARM (AMR)}
CHARM is an $N$-body+gasdynamics, AMR code, based on the CHOMBO-AMR library,
employing a higher order Godunov's method
for the solution of the hydrodynamic equations \citep{miniati06}.
Here a piecewice linear reconstruction scheme with Van Leer's limiter and
a nonlinear Riemann solver were used, resulting in a second order accurate
method in both space and time. CHARM was used to test the influence of IC on the cloud evolution. 

\subsubsection{Enzo (AMR)}
Enzo is an Eulerian AMR hybrid code (hydro + \emph{N}-body) code that was originally written by Greg Bryan and Michael Norman at the National Center for Supercomputing Applications at the University of Illinois \citep{bryan97}.
Enzo uses the Piecewise parabolic method \citep{collelawoodward:ppm} for solving fluid equations. PPM is a higher order accurate version of Godunov's method with an accurate piecewise parabolic interpolation and a non-linear Riemann solver for shock conditions. The method is third order accurate in space and second order in time. This together with the Riemann solver results in a very accurate shock treatment compared to the SPH codes where artificial viscosity is used.

\subsubsection{FLASH (AMR)}
FLASH is an AMR hybrid code (hydro + \emph{N}-body) developed by the ASC Center at the University of Chicago \citep{fryxel00}. The PPM hydrodynamical solver is formally accurate to second order in both space and time but performs the most critical steps to third- or fourth-order accuracy. For the simulations performed in this paper we have used FLASH version 2.3 using AMR with maximum refinement up to the resolutions indicated in Table \ref{table:simsummary}.

\subsubsection{Gasoline (SPH)}
Gasoline is a parallel Tree + SPH code, described in \cite{wadsley04}. The code is an extension to the N-Body gravity code PKDGRAV developed by \cite{stadel01}. Gasoline uses artificial viscosity (AV) to resolve shocks and has an implementation of the shear reduced version \citep{balsara95} of the standard \citep{monaghan92} artificial viscosity. Gasoline solves the energy equation using the asymmetric form
and conserves entropy closely. It uses the standard spline
form smoothing kernel with compact support for the softening of the
gravitational and SPH quantities.

The AV is implemented by solving a momentum equation of the form

\begin{eqnarray}
\frac{d\vec{v}_i}{dt}& = & -\sum_{j=1}^{n}m_j 
\left({\frac{P_i}{\rho_i^2}+\frac{P_j}{\rho_j^2}+\Pi_{ij}}\right)
\nabla_i W_{ij}, 
\label{eq:sphmom}
\end{eqnarray}
\noindent where $P_j$ is pressure, $\vec{v}_i$ velocity and the AV term $\Pi_{ij}$ is given by,
\begin{eqnarray}
\label{eq:artifvisc}
\Pi_{ij} = \left\{{ \begin{array}{ll}
\frac{-\alpha\frac{1}{2}(c_i+c_j)\mu_{ij}+\beta\mu_{ij}^2}{\frac{1}{2} (\rho_{i}+\rho{j})}
& {\rm for~}\vec{v}_{ij}\cdot\vec{r}_{ij}< 0,\\
~0 & {\rm otherwise}, \end{array}}\right.\\
{\rm\ where\ }
\mu_{ij} = \frac{h(\vec{v}_{ij}\cdot\vec{r}_{ij})}{\vec{r}_{ij}^{\,2}+0.01 (h_i+h_j)^2},
\label{eq:artifvisc2}
\end{eqnarray}
\noindent where $\vec{r}_{ij}=\vec{r}_i-\vec{r}_j$,
$\vec{v}_{ij}=\vec{v}_i-\vec{v}_j$ and $c_j$ is the sound speed. $\alpha$ and $\beta$ are the coefficients used for setting the viscosity strength, and are essential for capturing shocks and preventing particle interpenetration. Note that the viscosity term vanishes for non-approaching particles. The commonly used values in the literature is $\alpha=1$ and $\beta=2$ which originally was proposed by \cite{lattanzio86} using Sod shock tube tests. Later we will carry out experiments with different values of $\alpha$ and $\beta$.

\subsubsection{GADGET-2 (SPH)}
The TreeSPH code GADGET-2 \citep*{springel01,springel05} is the updated version of the GADGET-1. The code is similar in character to Gasoline but uses an entropy conserving formulation of SPH. This means that the thermodynamic state of each fluid element in GADGET-2 is defined through the specific entropy and not the specific thermal energy. GADGET-2 uses a somewhat different formulation of artificial viscosity than Gasoline. The viscosity term in Eq.\ref{eq:sphmom} is here formulated as
\begin{equation}
\Pi_{ij} =
-\frac{\alpha}{2} \frac{v_{ij}^{sig} w_{ij}}{\rho_{ij}} , 
\label{eq:ViscNew}
\end{equation}
where $v_{ij}^{sig}=c_{i} + c_{j} - 3 w_{ij}$ is the so called signal velocity. Here $w_{ij}=\vec{v}_{ij}\cdot\vec{r}_{ij}/|\vec{r}_{ij}|$ is the relative velocity projected onto the separation vector provided particles approach each other, otherwise the term vanishes just like in Gasoline.

\begin{table}
\caption{Simulation details. Enzo and ART use the static grids indicated in the table while the CHARM and FLASH simulations have been run using AMR up to the indicated resolution. All grid simulations are started from the 256,256,1024 initial conditions except for the analytically started CHARM run.}
\label{table:simsummary}
 \centering
 \begin{minipage}{140mm}
  \begin{tabular}{lcl}
\hline 
\hline
nParticles/Grid size & AV & Name \\ 
\hline
\hline
\textbf{ART}, static & &\\
64,64,256 & no AV & ART\_64 \\
128,128,512 & no AV & ART\_128 \\
256,256,1024 & no AV & ART\_256 \\
\hline
\textbf{CHARM}, AMR & &\\
512,512,2048 & no AV & CHARM\_512 \\
\hline
\textbf{Enzo}, static & &\\
64,64,256 & no AV & Enzo\_64 \\
128,128,512 & no AV & Enzo\_128 \\
256,256,1024 & no AV & Enzo\_256 \\
256,256,1024 & ZEUS, $\alpha=1$, $\beta=2$& Enzo\_{ZEUS} \\
\hline
\textbf{FLASH}, AMR & &\\
64,64,256 & no AV & FLASH\_64 \\
128,128,512 & no AV & FLASH\_128 \\
256,256,1024 & no AV & FLASH\_256 \\
\hline
\textbf{Gadget-2} & &\\
$10^7$ & $\alpha=0.8$& Gad\_10m \\ 
\hline  
\textbf{Gasoline} & &\\
$10^6$ & $\alpha=1$, $\beta=2$& Gas\_1m \\ 
$10^7$ & $\alpha=1$, $\beta=2$& Gas\_10m \\ 
$10^7$ & $\alpha=0$, $\beta=2.0$& Gas\_10mAV1 \\ 
$10^7$ & $\alpha=0$, $\beta=0.5$& Gas\_10mAV2 \\ 
$10^7$ & $\alpha=0$, $\beta=0.1$& Gas\_10mAV3 \\ 
$10^7$ & Balsara, $\alpha=1$, $\beta=2$& Gas\_Bals \\ 
\hline 
\hline
\end{tabular}
\end{minipage}
\end{table}

\section{Results of the simulations}
\begin{figure*}
\label{fig:denslices}
\begin{tabular}{cccc}
\psfig{file=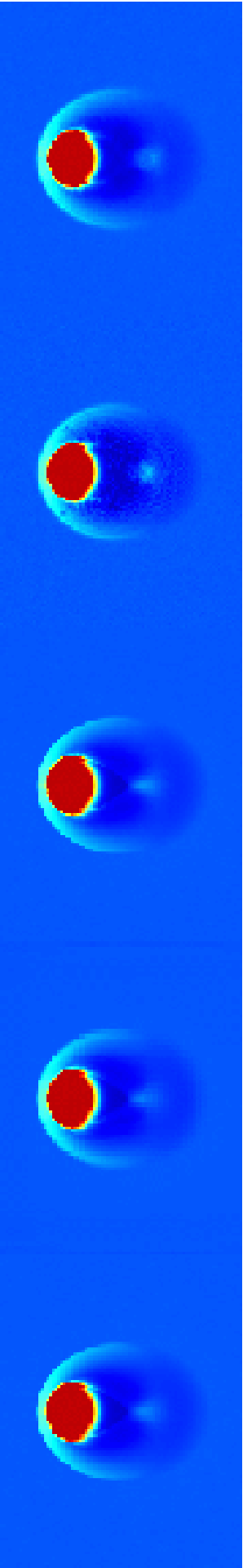,height=630pt} &
\psfig{file=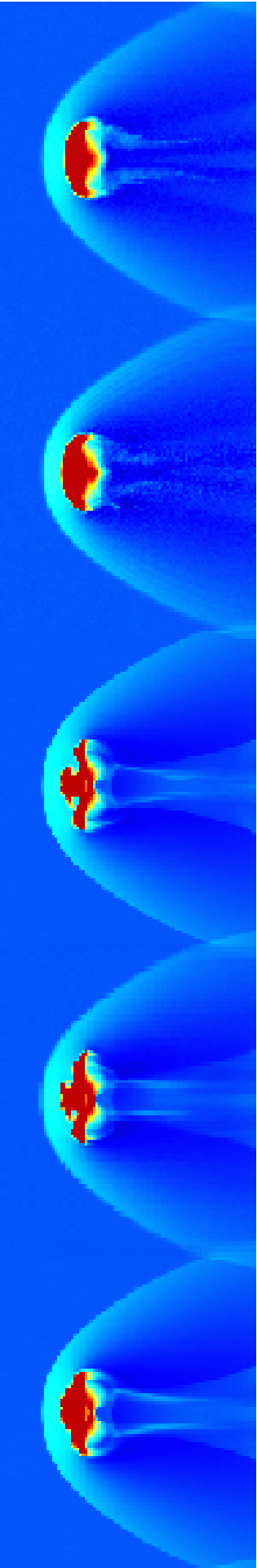,height=630pt} &
\psfig{file=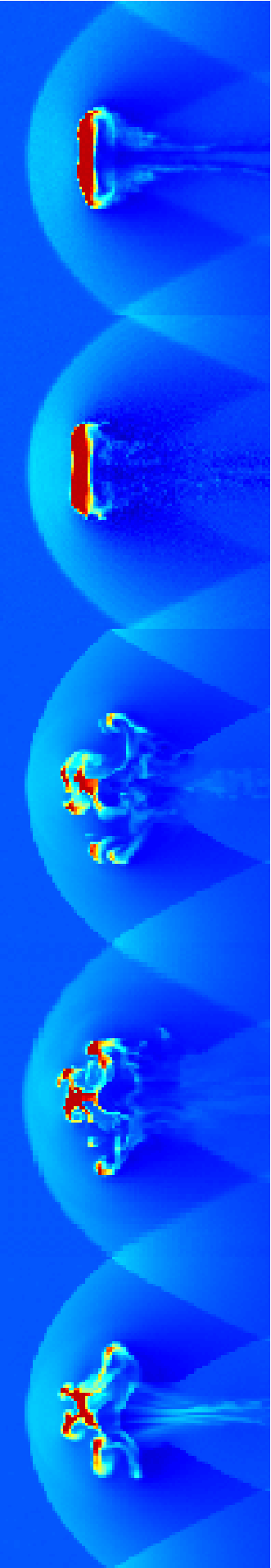,height=630pt} &
\psfig{file=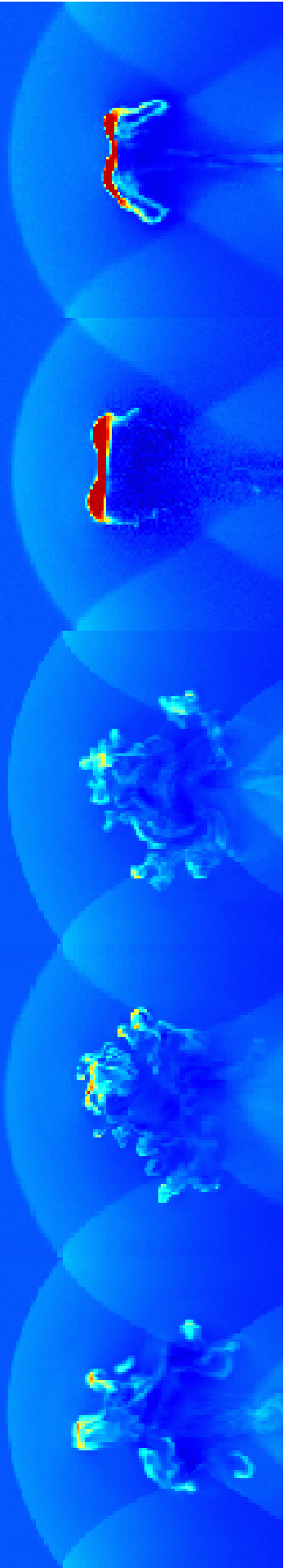,height=630pt} \\
\end{tabular}
\caption[]{Gas density slices through the centre of the cloud at $t=0.25,1.0,1.75$ and $2.5\, \tau_{\rm{KH}}$. From top to bottom we show Gasoline (Gas\_10m), GADGET-2 (Gad\_10m), Enzo (Enzo\_256), FLASH (FLASH\_256) and ART-Hydro (ART\_256). The grid simulations clearly show dynamical instabilities and complete fragmentation after $2.5\, \tau_{\rm{KH}}$, unlike the SPH simulations in which most of the gas remains in a single cold dense blob.} 
\end{figure*}

Fig. 4 shows central density slices of Gasoline (Gas\_10m), GADGET-2 (Gad\_10m), Enzo (Enzo\_256), FLASH (FLASH\_256) and ART-Hydro (ART\_256). These are the high resolution simulations with the default standard settings.

\begin{figure*}
\label{fig:SPHslices} 
\begin{tabular}{cc}
\psfig{file=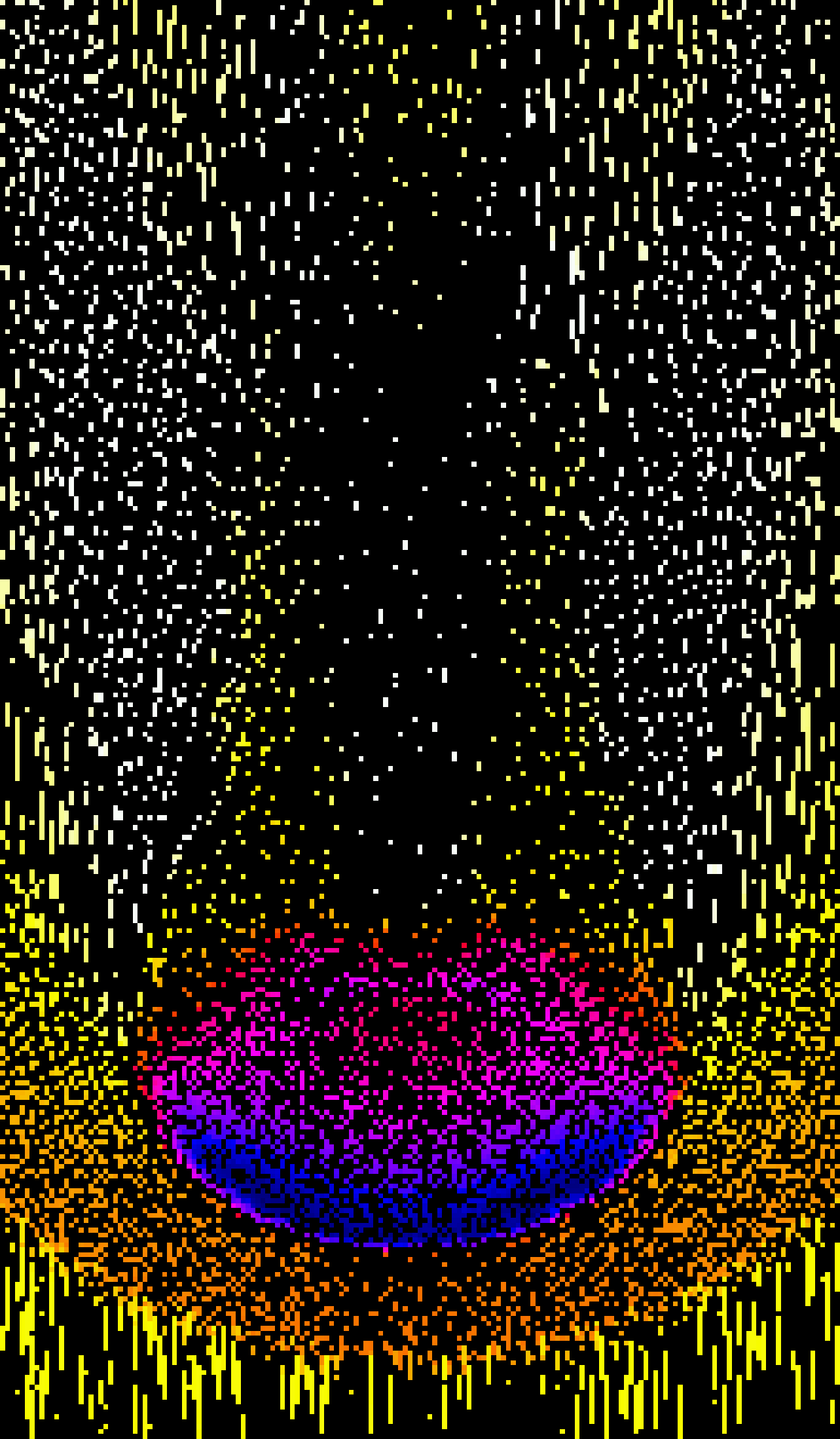,height=290pt} &
\psfig{file=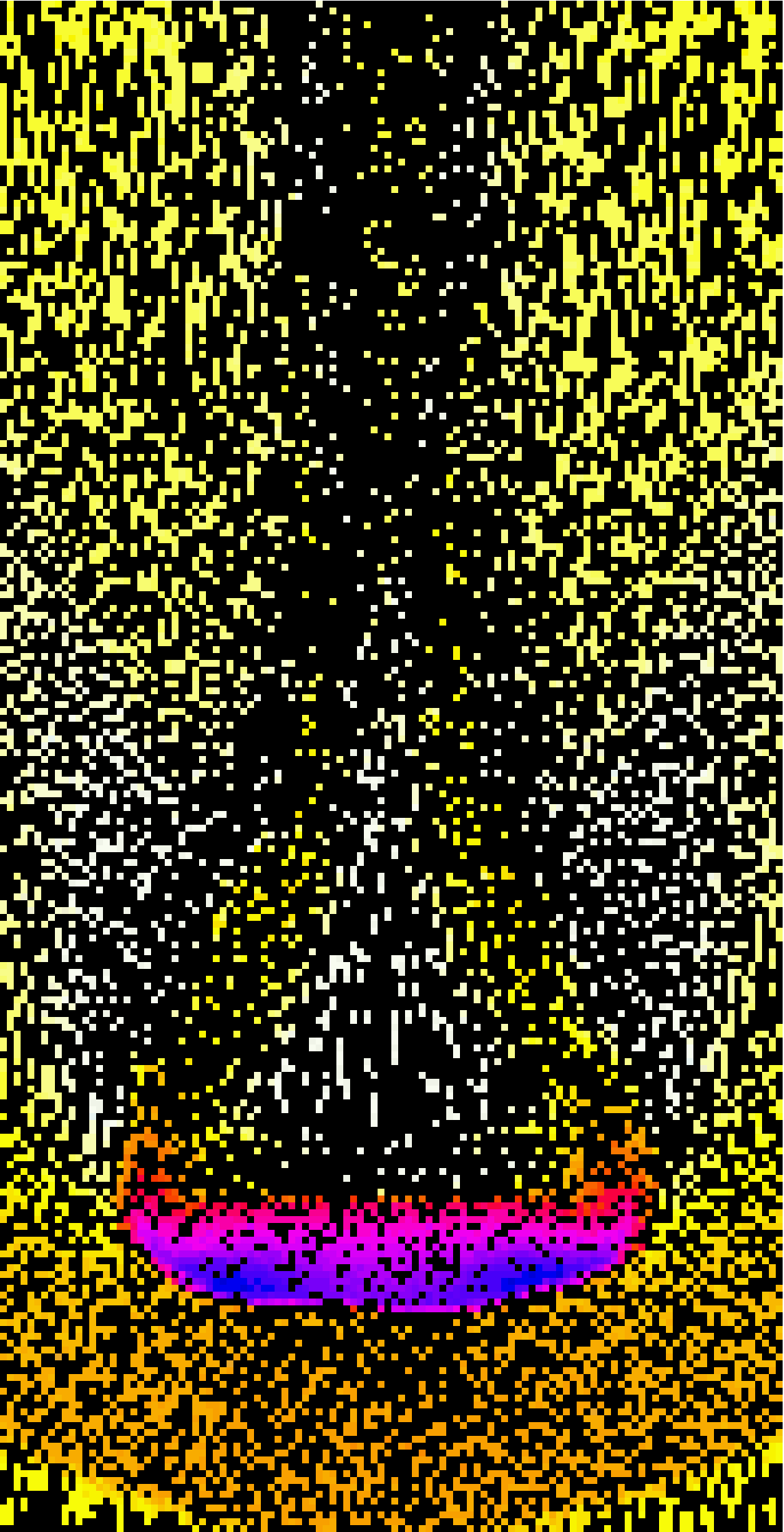,height=290pt} \\ 
\end{tabular}
\caption[]{A thin central slice of the SPH particles of Gas\_10m at $t=0.75\,\tau_{\rm{KH}}$(left panel) 
and $t=1.5\,\tau_{\rm{KH}}$ (right panel). The density ranges from high (blue) to low (white) and the magnitude of the velocity vectors are normalized to the reference frame of the centre of the cloud. We clearly see the effect of the cloud stretching due to the lateral Bernoulli zones and the formation of downstream vorticity. } 
\end{figure*}

The simulations of the two SPH codes, Gas\_10m and Gad\_10m, show a very similar evolution. As expected, a detached bow shock forms directly in front of the cloud. An internal shockwave forms within the cloud compressing it. The post shock flow encompasses the cloud, creating Bernouilli zones on the top and bottom with lower pressure. This causes the cloud to become elongated as well as compressed along the $z$-axis and we see gas being ablated, i.e. stripped through the induced pressure differences, from the top and bottom edges. Gas stripping slowly progresses and the cloud's shape does not change significantly for a long time \citep{doroshkevich81}. Fig. \,\ref{fig:SPHslices} shows the particles in a thin slice centred on the cloud. The velocity vectors of each particle are plotted in a reference frame centred on the cloud. The colours indicate the gas density. Behind the edges of the cloud we see a vortex created due to the shearing motion of the ambient medium which creates a low pressure region behind the cloud.
Initially, the cloud evolution is similar in the grid simulations. It is compressed and elongated and gas is removed from the trailing edges where the vortex has created a vacuum behind the cloud. Some of the ambient medium turns around and falls onto the backside of the cloud.
However, the late cloud evolution is quite different in these simulations. Early on we observe surface perturbations on the front of the cloud, probably originating from the way the ICs are setup (see argument in sect.\,\ref{sect:IC}). A complicated mixture of KHIs and RTIs are developing on the cloud front which, due to subsequent compression and lateral expansion, becomes even more KH and RT unstable. By  $t\sim\tau_{\rm{KH}}$, large scale KHIs have developed and the cloud starts to fragment. Further instabilities and turbulence mixes the smaller clumps of gas into the ambient medium.  All grid simulations show basically the same cloud destruction time. We also note that Eulerian (shock capturing) methods effectively localise shocks to a few grid cells compared to the smoothed out shocks in the SPH simulations resulting from AV shock capturing schemes.

\begin{figure}
\psfig{file=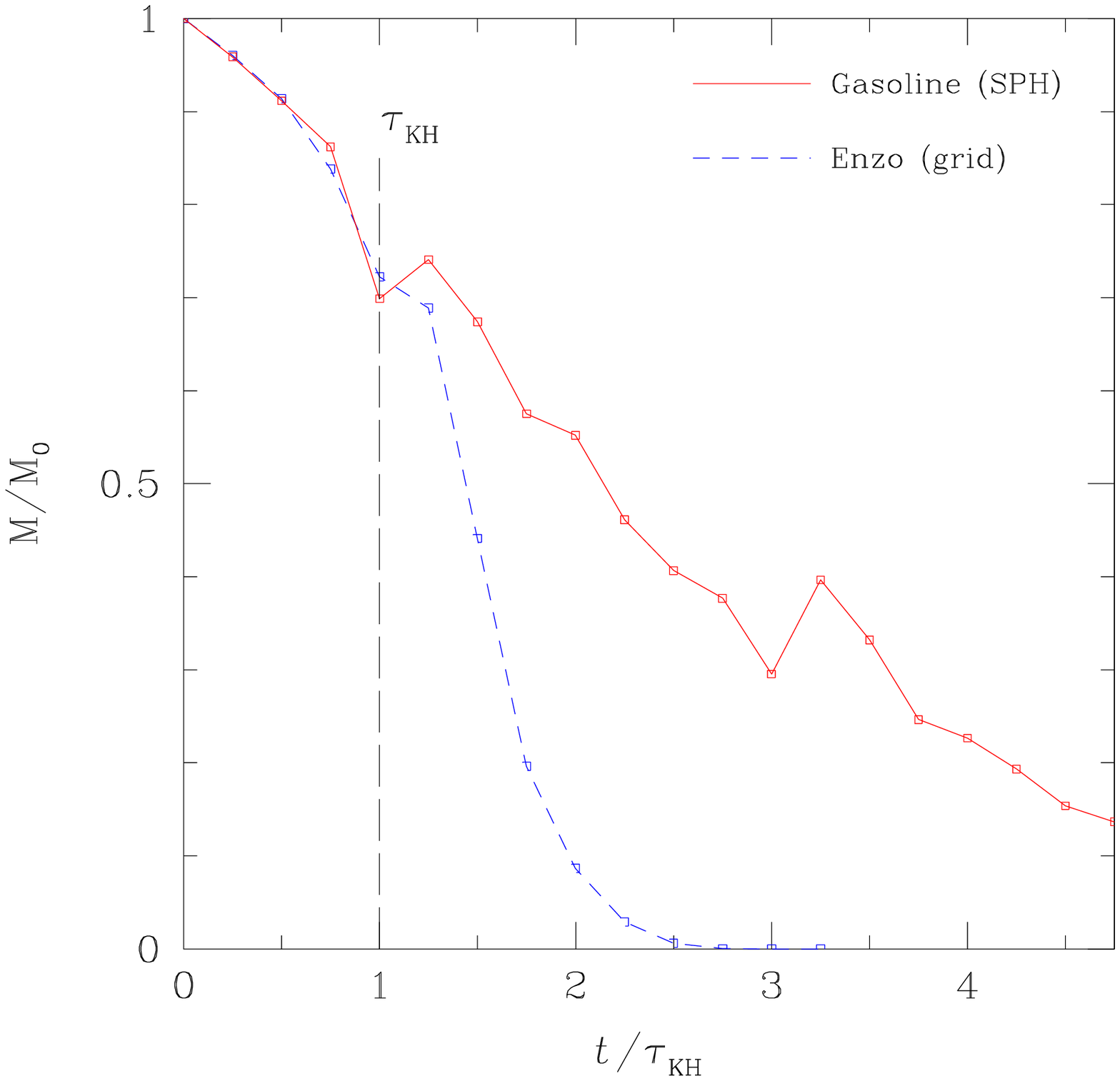,width=240pt,bbllx=30bp,bblly=184bp,bburx=582bp,bbury=708bp,clip=}
\caption[]{The evolution of the cloud mass fraction. In the SPH simulation (solid, red), the cloud slowly loses mass to the ambient medium and has not been completely mixed even after $5\,\tau_{\rm{KH}}$. The grid simulation (dashed, blue) follows the SPH up to the time at which the KH instability causes it to rapidly 
fragment and mix.}
\label{fig:massfrac}
\end{figure}

In Fig. \,\ref{fig:massfrac} we show the remaining cloud mass fractions as a function of time for the Enzo and Gasoline simulations. These are representative of grid and SPH methods. We define the cloud as being any gas that satisfies $T < 0.9\,T_{\rm{ext}}$ and $\rho > 0.64\,\rho_{\rm{cl}}$. It is of course possible to construct more elaborate criteria but these select the gas that visually is a part of the cloud. The figure shows that both techniques give a similar mass loss up to $\sim\tau_{\rm{KH}}$. Before this time the gas loss is mainly due to ablation into the low pressure zone created behind the cloud. 
As soon as we pass $\tau_{\rm{KH}}$ for large scale KHI the SPH and grid methods diverge. In the grid simulation, the cloud quickly disrupts and diffuses into the ambient medium, while the SPH simulation only shows continuing stripping. After $t=2.5\tau_{\rm KH}$, no gas in the grid simulation can satisfy our criteria while the SPH simulation still shows a mass fraction of $\approx 40\%$. This shows us that the vortex shedding through the Bernoulli zones is the most important mechanism for mass-loss at $t<\tau_{\rm{KH}}$ in both methods. After this time dynamical instabilities dominate the grid mass-loss.

\subsection{Resolution dependence}
\begin{figure*}
\begin{tabular}{cccc}
\psfig{file=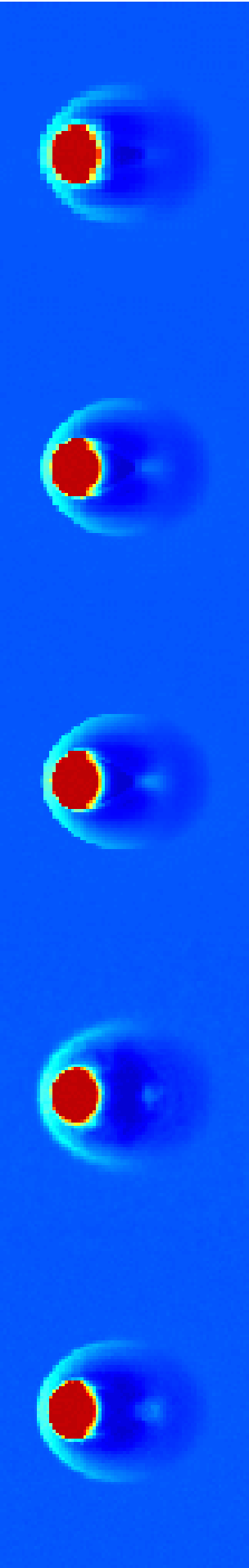,height=630pt} &
\psfig{file=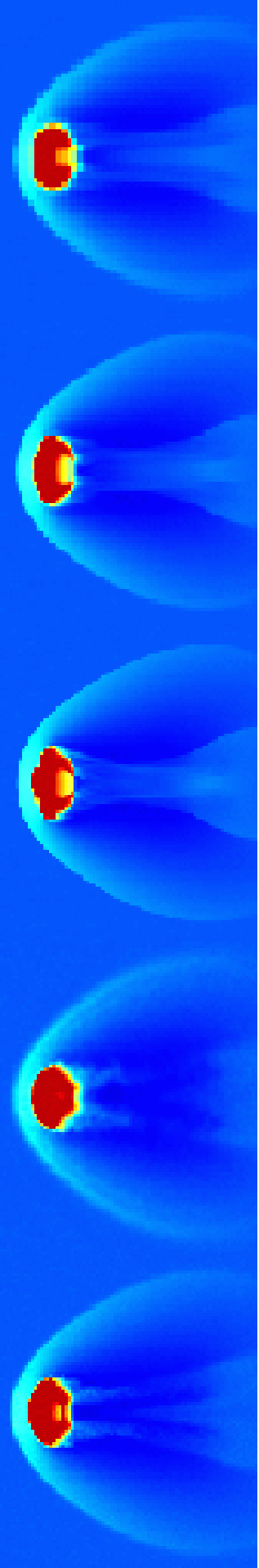,height=630pt} &
\psfig{file=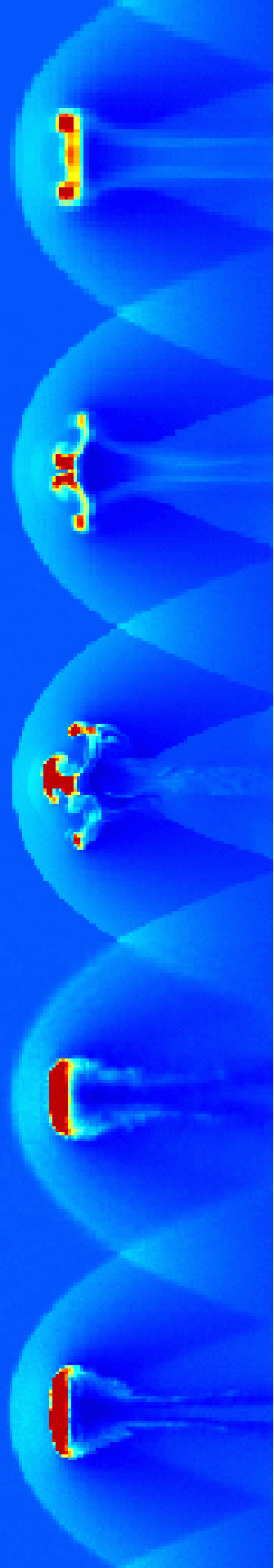,height=630pt} &
\psfig{file=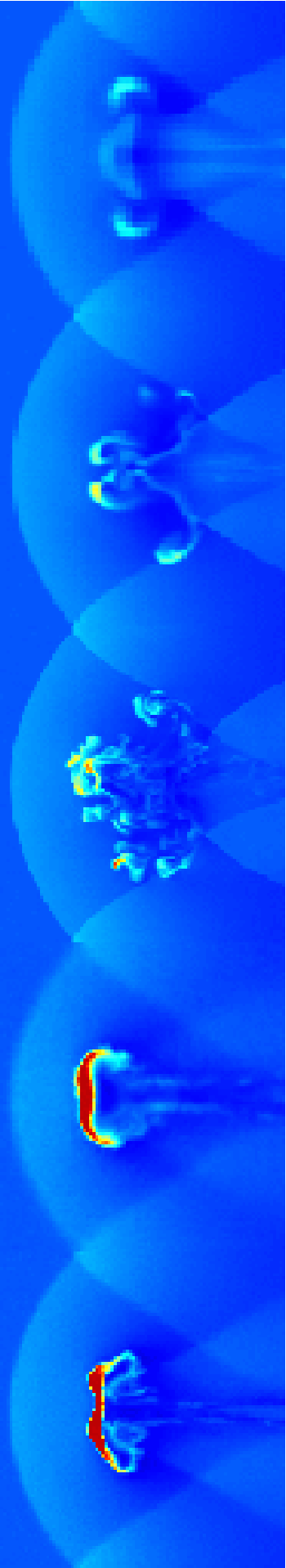,height=630pt} \\
\end{tabular}
\caption[]{Resolution study for Enzo and Gasoline. The panels show density slices of, from top to bottom, Enzo\_64, Enzo\_128, Enzo\_256, Gas\_1m and Gas\_10m for $t=0.25,0.75,1.5$ and $2.25\,\tau_{\rm{KH}}$. We see that resolution changes the phase of the instabilities in the grid simulations while the destruction time is the same. Higher resolution also shows less diffusion and better resolves small scale fragments. The Gasoline runs are not able to resolve small scale instabilities at all. } 
\label{fig:resolution}
\end{figure*}
It is difficult to do a direct translation between grid and SPH resolution. The maximum allowed resolution is a fixed grid of size 256x256x1024 in the grid runs and $10^7$ particles in the SPH runs. This means that there is almost a factor 7 more cells compared to particles. On the other hand, cells are uniformly distributed in space and only $\approx 21378$ cells cover the cloud, assuming an IC setup. This should be compared to the $100,000$ particles constituting the cloud in the high resolution SPH run. A comparison like this is still not straightforward due to the fact that SPH uses particles as \emph{non-independant} resolution elements. This means that each particle is not a carrier of information without neighbours to smooth over, and the effective number of resolution elements is more or less set by the kernel shape and number of neighbours to smooth over.

Resolution affects the convergence of hydrodynamical simulations. A cut-off is always introduced on the scale of the spatial resolution below which instabilities can not be resolved. This often serves as a source of numerical viscosity.

For most of the codes used in the comparison, we have varied the resolution in order to obtain an understanding of how this changes the cloud morphology, mass loss and fragmentation time (see Table\,\ref{table:simsummary}). Fig. \,\ref{fig:resolution} shows the outcome of, from top to bottom, Enzo\_64,  Enzo\_128,  Enzo\_256,  Gas\_1m,  and Gas\_10m. In the grid simulations we conclude that, while the compression and elongation of the cloud is relatively similar, the detailed way the cloud fragments is resolution dependent, owing to the differences in IC \citep{jones96}. In Enzo\_64, a phase of the KHI centered on the clouds symmetry axis is dominant. This phase is less pronounces as resolution is increased (Enzo\_128) and it is nowhere to be seen in In Enzo\_256. Going to higher resolution we see more and more small scale instabilities developing which enhance mixing of the cloud material with the background flow. Numerical diffusion is stronger in low resolution simulations which is why parts of the cloud survive longer in the higher resolution runs.

The different SPH simulations are qualitatively very similar. Instabilities can not be resolved in Gas\_1m nor in Gas\_10m. However, we note a weak large scale RT instability on the cloud front at $t=2.25\,\tau_{\rm{KH}}$ in Gas\_10m, which is absent in Gas\_1m.

\begin{figure}
\psfig{file=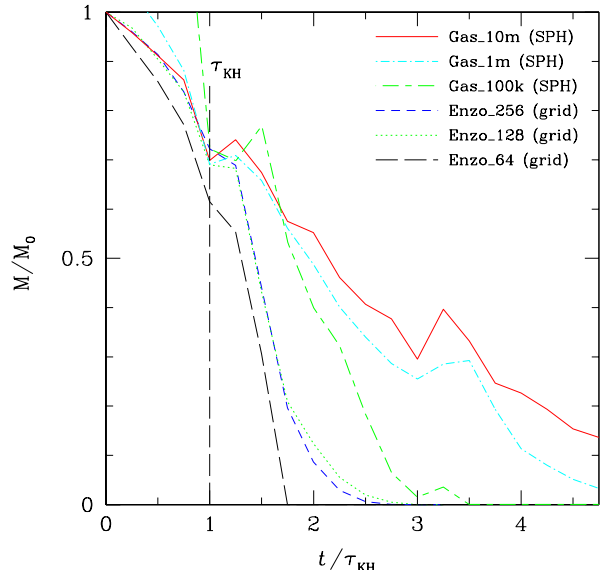,width=240pt,bbllx=30bp,bblly=184bp,bburx=582bp,bbury=708bp,clip=}
\caption[]{The evolution of the cloud mass fraction for different resolutions. 
As the resolution of the grid simulations is increased from 64 to 128 to 256 cells across the wind tunnel, the amount of mass increases a little but converges. Increasing the resolution of the SPH simulations does not decrease the amount of mass lost, rather the opposite, perhaps due to the momentum transfer due to massive particles acting like "`bullets"'.}
\label{fig:massfracres}
\end{figure}

The general description above is again quantified by studying the cloud mass fraction at each timestep, see Fig.\,\ref{fig:massfracres}. In this plot we have also added an extra low resolution SPH simulation using only $100\,000$ particles. The grid simulations show a clear trend of dissolving the cloud quickly after $\sim\tau_{\rm{KH}}$ regardless of resolution while the SPH simulations only show a steady mass loss due to the material ablated into the trailing vacuum. Decreasing the SPH resolution causes the mass fraction to rise above the initial value during the initial phase and mass is lost more rapidly for $t>\tau_{\rm{KH}}$. The latter effect is most probably due to the increased mass of each particle, causing each particle interaction to transfer momentum in a more violent, ``bullet-like'' fashion.
\subsection{Initial Seeds}
\begin{figure*}
\begin{tabular}{cccc}
\psfig{file=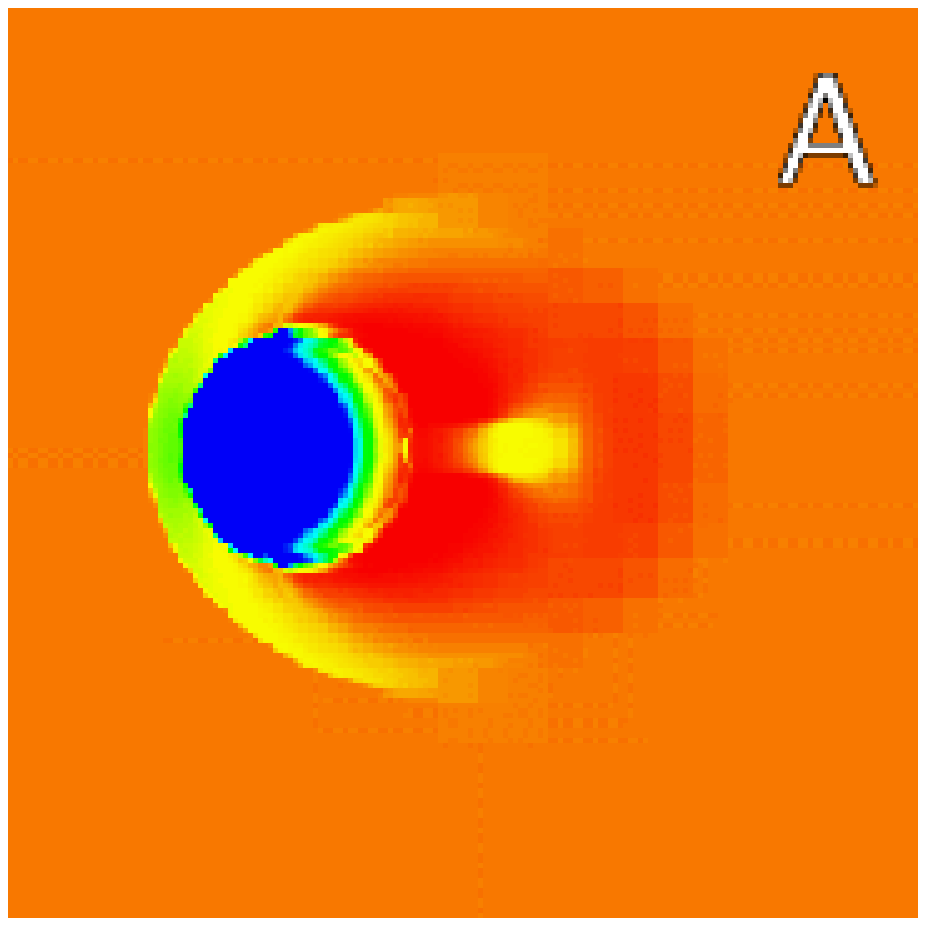,width=115pt} &
\psfig{file=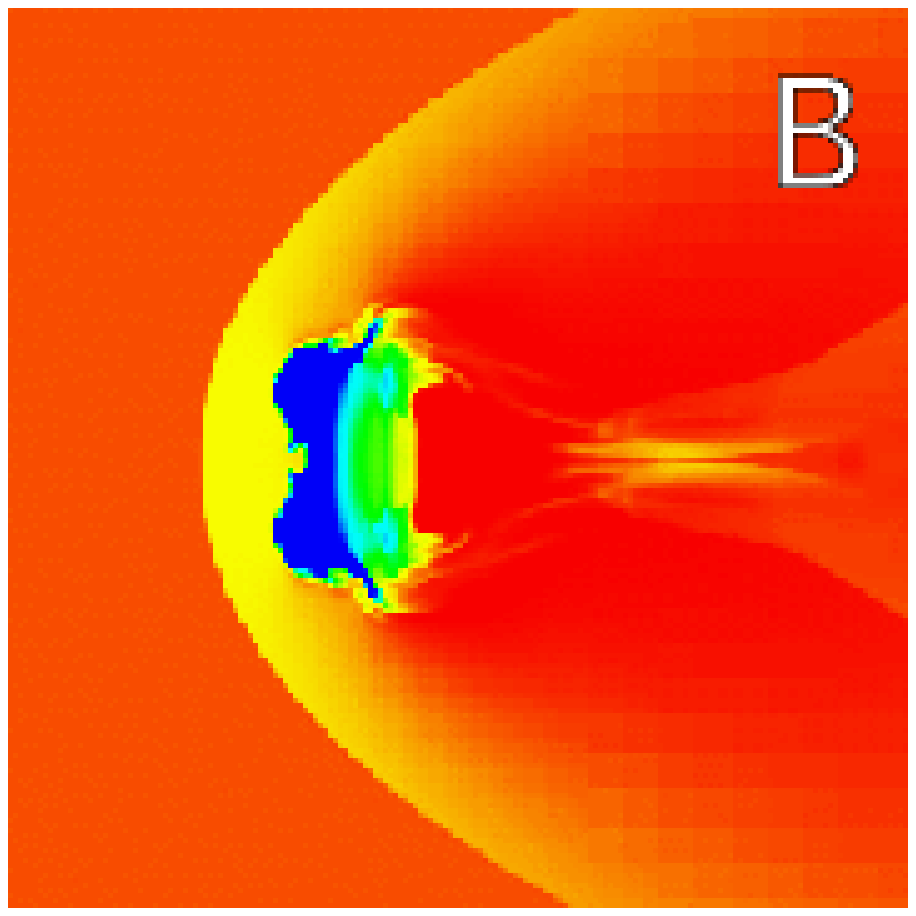,width=115pt} &
\psfig{file=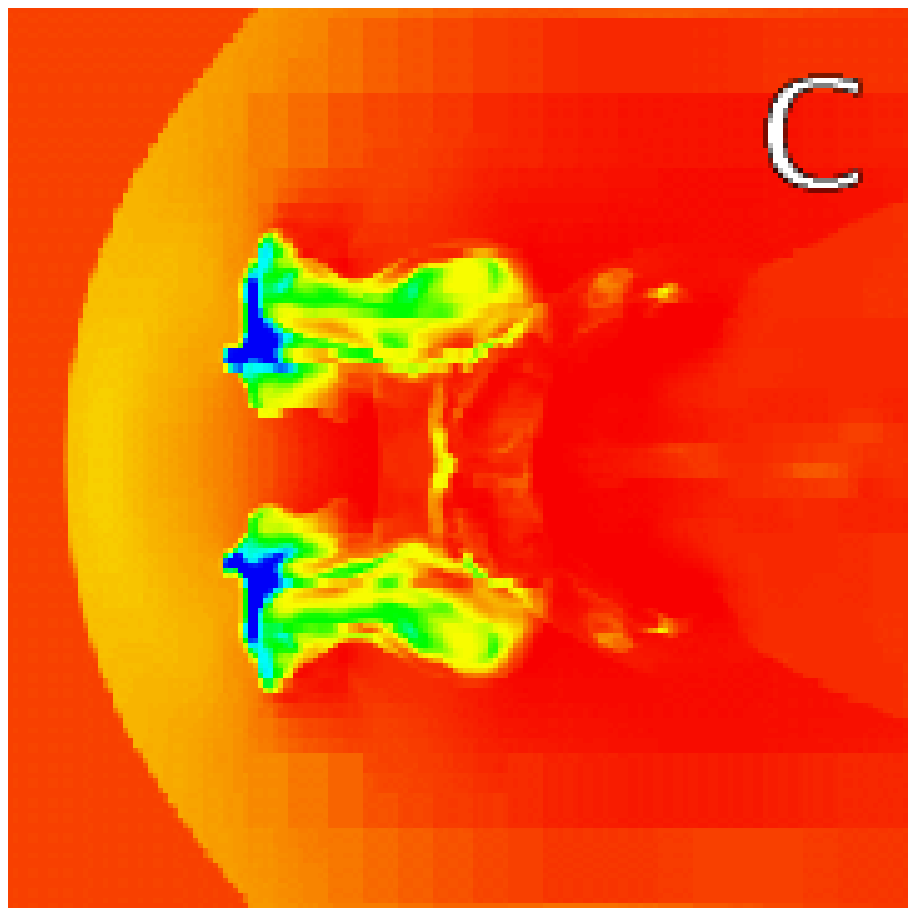,width=115pt} &
\psfig{file=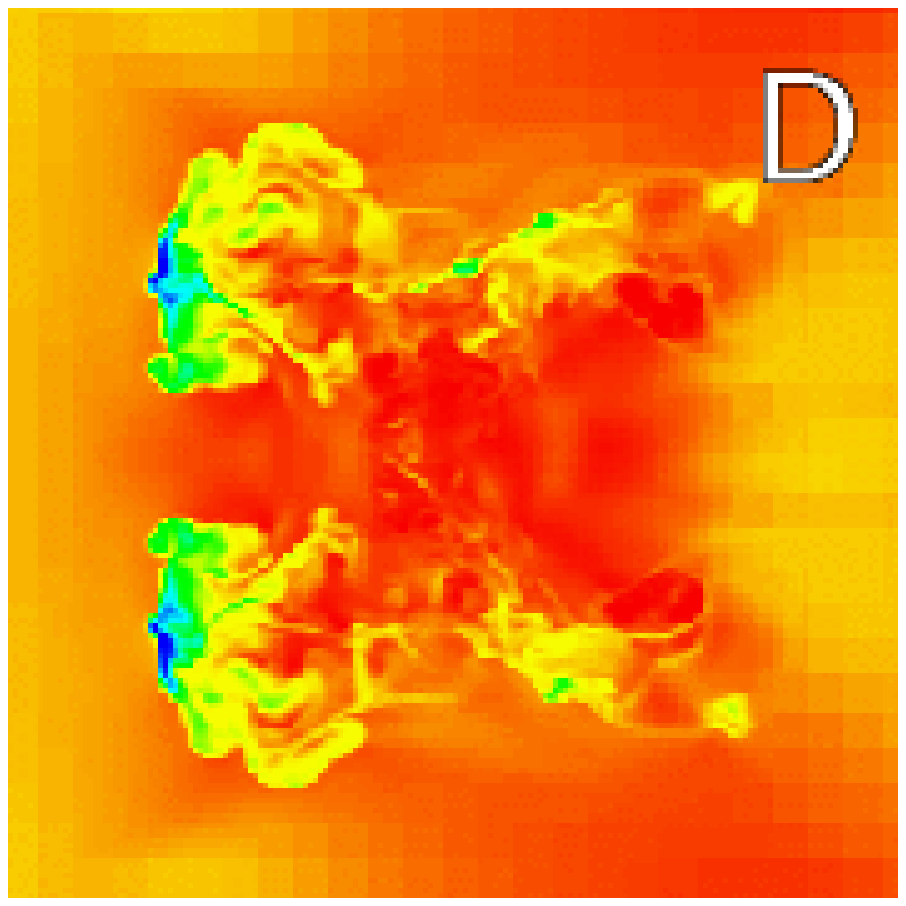,width=115pt} \\
\end{tabular}
\caption[]{Evolution of the cloud with `analytic' initial conditions. 
Each frame shows a density slice through the
cloud center at times $t=0.24, 0.9, 1.7$ and $2.5\,\tau_{\rm{KH}}$ with densities varying from low (red) to high (blue).}
\label{fig:charmfig}
\end{figure*}

As partly shown in the previous test, the development of the
instabilities, particularly during the nonlinear stages, is sensitive
to the exact definition of the initial conditions.  This is because
they set the seed perturbations out of which the instabilities grow.
However, while the mixing of the cloud material with the background
medium is affected by small scale motions that arise from the small
unstable scales, the cloud disruption is mostly the result of the
development of the large scale perturbations.

As an example of this in Fig.~\ref{fig:charmfig}
we show the evolution of the cloud-wind interaction but
with initial conditions set directly from the analytic definition.
Thus in this case the initial conditions are free of noise and are
purely symmetric. A base grid of $(32\times 128 \times128)$ was used
with two additional levels of refinement with refinement ratio of 4
placed dynamically in regions where the relative change in density,
$\Delta \rho/\rho$  exceeded 20\%. This corresponds to an effective
resolution of $512\times 512\times 2048$ in the finest grids,
which reduces the level of perturbation with respect to the previous
cases.

As shown in panel B of Fig. \,\ref{fig:charmfig} the most
destructive mode has a different phase than in the cases illustrated
above for the corresponding grid based codes.
However, as in the previuos cases,
by $t=2.5\,\tau_{\rm{KH}}$ (panel D) the cloud has been completely 
reduced to
debris by the instabilities. This shows that despite differences
in the appearence of the cloud gas distribution its fundamental
fate of disruption and subsequent mixing on a timescale of a few 
$\tau_{\rm{KH}}$ is independent of the specific definition of the 
initial conditions.  

\section{Why so different?}
What is the reason for the observed discrepancies between simulations carried out using SPH and grid-based techniques?
Differences between SPH and grid-based results have been discussed before in the literature 
\citep{pearce99,thacker00,ritchie01,tittley01,springel02,marri03} in different contexts to this study.
While artificial viscosity is the most obvious focus for criticism of SPH it is not the main reason for 
the differences observed in this test. We will show this in section~\ref{sect:AV} before focusing on the 
almost complete suppression of KH (and RT) instabilities in SPH simulations of this test and present an 
explanation of why this occurs.

\subsection{Artificial viscosity}
\label{sect:AV}
\begin{figure*}
\begin{tabular}{cccc}
\psfig{file=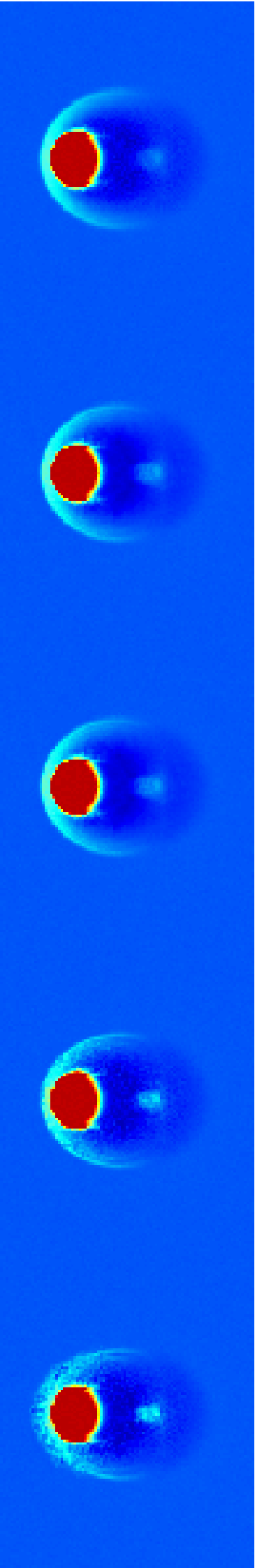,height=630pt} &
\psfig{file=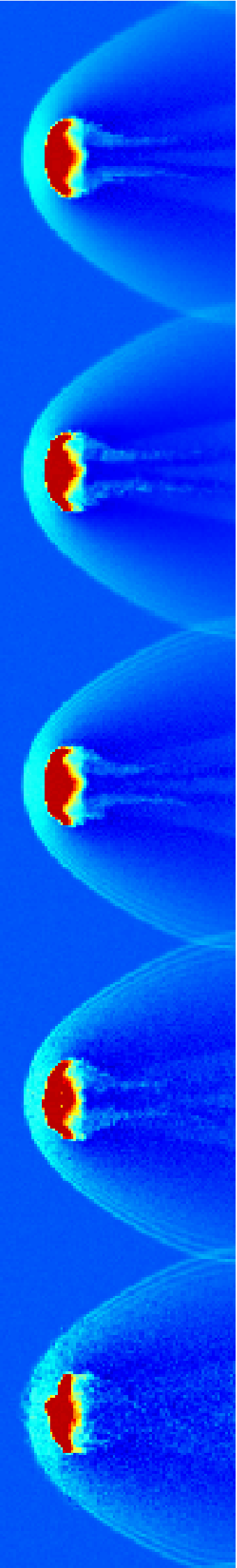,height=630pt} &
\psfig{file=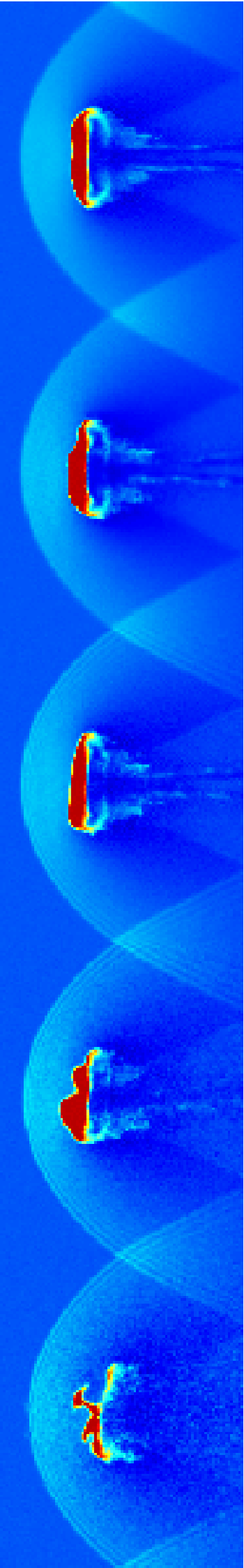,height=630pt} &
\psfig{file=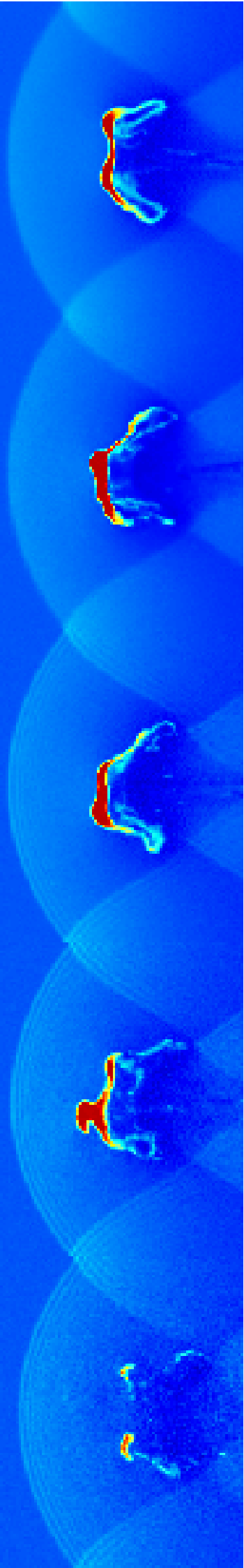,height=630pt} \\
\end{tabular}
\caption[]{Viscosity study for Gasoline. The panels show density slices of, from top to bottom, Gas\_10m, Gas\_Bals, Gas\_10mAV1, Gas\_10mAV2 and Gas\_10mAV3 for $t=0.25,0.75,1.5$ and $2.25\,\tau_{\rm{KH}}$. We can see how reducing shear viscosity and removing the bulk viscosity renders very similar results; the cloud destabilizes to a higher degree. By reducing the shock capturing viscosity the cloud destabilizes even further, most probably to an unphysical solution in the lower setting. The artificial post shock ringings also gets more pronounced, as expected for lower viscosity settings. } 
\label{fig:AV}
\end{figure*}

The artificial viscosity $\beta$ parameter in Eq.\,\ref{eq:artifvisc} is necessary for shock capturing and is required for SPH to work properly in supersonic regimes. The $\alpha$ parameter has a less obvious meaning and the classical $\alpha=1.0$ setting is most probably unphysical. It can be argued \citep[e.g.][]{watkins96} that $\alpha$ can roughly be interpreted as a Navier-Stokes shear plus bulk viscosity, even though the AV is only sensetive to flow properties such as interparticle travelling. Bulk viscosity is normally not important in fluid dynamics, except in the theory of attenuation of sound waves \citep[e.g.][]{faber95}. In numerical simulation its inclusion is for the most part to dampen the so called \emph{post-shock ringing}.

The inclusion of artificial viscosity leads us to one of the first possibilities for the observed discrepancy: \emph{We are not solving the same hydro-dynamical equations in the different codes}. By adding AV we are solving some kind of Navier-Stokes equation when we actually want to compare the solutions to the grid codes that, in this sense, are closer to the Euler equations. Note however that there is always numerical viscosity due to resolution and truncation in \emph{all} simulation methods.

Viscosity has two major effects on the processes we want to capture in this test:
\begin{enumerate}
\item Dampening of small scale velocity perturbations and random velocities
\item Diffusion of post shock vorticity and thus smearing of turbulence
\end{enumerate}
The effect of (i) will enter as a stabilizing factor for the growth of instabilities. Physical kinematic viscosity, $\nu$, sets a cut off for the size of the smallest eddies in turbulence \citep{shu92}, below which turbulent motion is diffused. 

The effect of (ii) follows from the first one and is obvious from inspection of the vorticity transport equation \citep[e.g.][]{shu92} 
\begin{equation}
\frac{\partial \mbox{\boldmath$\omega$}}{\partial t} + 
\nabla \mbox{\boldmath$\times$}
\left(\mbox{\boldmath$\omega \times v$}\right) = \nabla P \times 
\nabla \left(\frac{1}{\rho}\right)  +\nu \nabla^2\mbox{\boldmath$\omega$}   \, ,
\label{eq:vorticity}
\end{equation}
where $\mbox{\boldmath$\omega$}\equiv\nabla\times\mbox{\boldmath$v$}$ is the vorticity. The two terms on the right hand side can create or diffuse vorticity. The first term is the \emph{baroclinic} term which is non vanishing if we have non-aligned pressure and density gradients. This is the case in oblique shocks like in the bow shock of our cloud simulation. The second term is responsible for diffusing vorticity in space i.e. taking local vorticity and spreading it into the general flow. This means that as soon as we have viscosity, \emph{we will dampen vorticity}. Especially important is the vorticity generated in the post shock flow, which should act to destabilize the cloud together with the surface instabilities. 

A study on how artificial viscosity dampens small scale vorticity was made by \cite{dolag05}. By using a low viscosity formulation of SPH they find higher levels of turbulent gas motions in the ICM and noted that shocked clouds tend to be unstable at earlier times. However, by looking at their Figure 3 we note that the overall difference in the cloud evolution is small.

In order to understand the effect of artificial viscosity in our cloud-wind test we have performed three simulations with modified setting of the viscosity coefficients. These are Gas\_10mAV1, Gas\_10mAV2 and Gas\_10AV3, see Table \,\ref{table:simsummary} for viscosity values. A simulation using the Balsara switch but with the standard ($\alpha=1.0$, $\beta=2.0$) was also performed. Fig. \,\ref{fig:AV} shows the outcome of the simulations at $t=0.25, 0.75, 1.5$ and $2.25\tau_{\rm{KH}}$. We can directly see the strong impact these terms have on the stability of the simulation. The standard $\alpha=1.0$, $\beta=2.0$ is the most stable one, most probably due to the unphysical use of the $\alpha$ bulk viscosity. The use of $\alpha=0$ and $\beta=2.0$ or the Balsara switch renders very similar visual results. This is because the Balsara switch turns of viscosity where  $\nabla$\mbox{\boldmath$\times v$} is significant, which is the case for shearing flows like on the surface of the cloud. Note that $\nabla$\mbox{\boldmath$\times v$} is a very noisy quantity when measured using only 32 neighbours. By further lowering the shock capturing $\beta$ viscosity we make the cloud even more unstable but it is not clear how physical this solution is. The shock front gets more blurred and we see strong post shock ringing effects. The reason for the increased instability in the $\alpha=0$, $\beta=0.5$, and $\alpha=0$, $\beta=0.1$ case is most probably due to high speed particles traveling through the poorly captured shock region and transferring momentum inside the cloud, perturbing it in an unphysical way.

We see from these simulations how lowering viscosity will make the cloud less stable, which is expected from linear analysis. However, we still can't obtain agreement with the grid based codes. This leads us to suspect that there are more fundamental reasons behind the discrepancies.

\subsection{Resolving instabilities}
\begin{table}
\caption{Performed KH runs}
\begin{tabular}{lccccl}
\hline 
\hline
Resolution & $\chi$ & $\delta v/v_{\rm shear}$ & $\tau_{\rm{KH}} $ & IC & Name \\ 
\hline
\hline 
 \textbf{Enzo} & & & & & \\
$\{256,256,8\}$ & 8.0 & $1/80$ & $1.70$ & lattice & GRID1 \\
$\{256,256,8\}$ & 10.0 & $1/40$ & $1.86$ & poisson & GRID2 \\
$\{256,256,8\}$ & 10.0 & $1/40$ & $1.86$ & glass & GRID3 \\
\hline 
 \textbf{Gasoline} & & & & &\\
$900\,$k part & 8.0 & $1/80$& $1.70$ & lattice & SPH1 \\
$1.1\,$M part & 10.0 & $1/40$& $1.86$ & poisson & SPH2 \\
$1.1\,$M part & 10.0 & $1/40$& $1.86$ & glass & SPH3 \\
\hline 
\hline
\label{table:KH}
\end{tabular}
\end{table}

In order to create an even simpler test problem to compare instabilities between codes, we carried out a classical Kelvin-Helmholtz test using Gasoline and Enzo. We looked at the shearing motion of two gases of different densities and with small perturbations imprinted at the boundary. This captures some of the hydrodynamics at the surface of the cloud in the blob test.

The setup is a periodic box with dimensions $\{L_x.L_y,L_z\}=\{1,1,1/32\}$, divided into two regions: one cold, high density and one warm, low density. The density and temperature ratio is $\chi=\rho_{\rm b}/\rho_{\rm t}=T_{\rm t}/T_{\rm b}=c_{\rm t}^2/c_{\rm b}^2$, putting the whole system in pressure equilibrium. The two layers are given constant and opposing shearing velocities, with the top layer moving leftward at a Mach number $\mathcal{M}_{\rm{t}}=v_{\rm t}/c_{\rm t} \approx 0.11$ and the bottom layer moving rightward at a Mach number $\mathcal{M}_{\rm{b}}=v_{\rm b}/c_{\rm b} \approx 0.34$ in the case of $\chi=10$. The shear velocity becomes $v_{\rm shear}=0.68\,c_{\rm b}$ and the subsonic regime will assure growth of instabilities \citep{vietri97}. This setup should mimic the growth of instabilities on the cloud surface.
 
To trigger instabilities we have imposed sinusoidal perturbation on the vertical velocity of the form
\begin{equation}
\label{eq:perturbation}
v_y(x) =\delta v_{\rm y}\sin(\lambda 2\pi x),
\end{equation}
where  $\delta v_{\rm y}$ is the amplitude of the perturbation in terms of the sound speed $c_{\rm{b}}$ and $\lambda$ is the wavelength of the mode which we have put to $1/6$ in all of out tests. The perturbation is limited to a central strip around the interface of thickness $5\%$ of the box size. 

The initial conditions are again first generated using particles which are then mapped to the grid so that both codes have very similar starting points. An important issue for this type of test is how particles are distributed since this will introduce a certain amount of noise via discreteness. The most common techniques for this are:
\begin{itemize}
\item {\bf Lattice:} Particles ordered in a perfect grid. This minimizes local density fluctuations. This type of IC is optimal for grid codes.
\item {\bf Poisson:} Particles are randomly distributed which generates local density variations, causing spurious pressure forces.
\item {\bf Glass:} Particles are heated and relaxed until equilibrium and homogeneity is found. This type of simulated annealing will create a relaxed system and is more optimal for SPH than for grids. 
\end{itemize}
Any initial condition that has local density variations will trigger small scale KH instabilities. We carried out this test using all three methods in order to illustrate their impact. The lattice is obviously perfect for grid codes, making a perfectly homogeneous gas. 
This quality does not automatically produce clean SPH initial conditions due to the averaging over nearby particles. The poisson ICs are very noisy in both the grid and SPH case, even though grid codes tend to smooth the noise over the cell sizes. The glass IC is intuitively the closest IC for both methods producing a self-consistent and homogeneous initial state for SPH simulations while leaving only small fluctuations for both grid and SPH methods.  

\begin{figure*}
\begin{tabular}{ccc}
\psfig{file=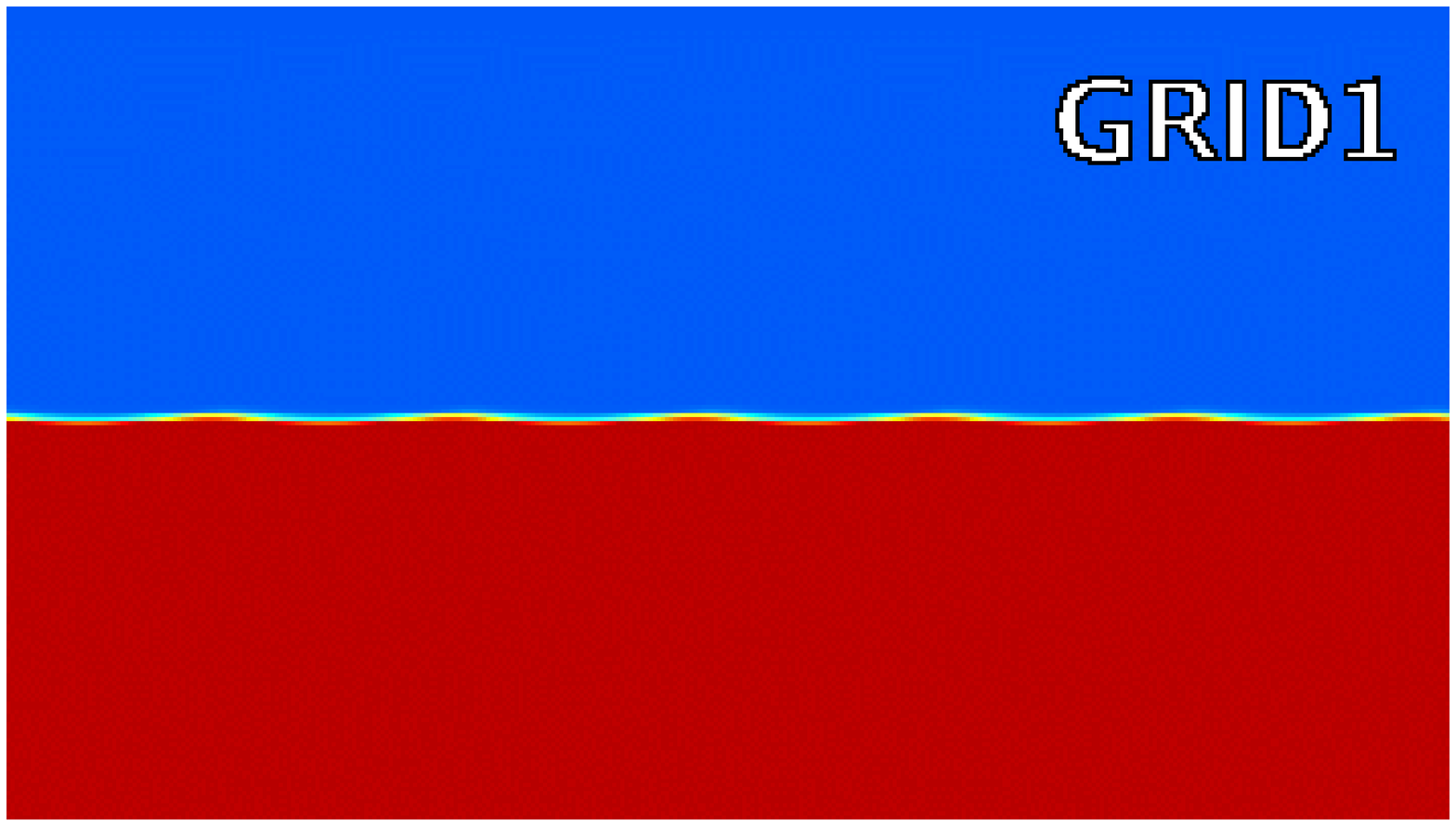,width=160pt} &
\psfig{file=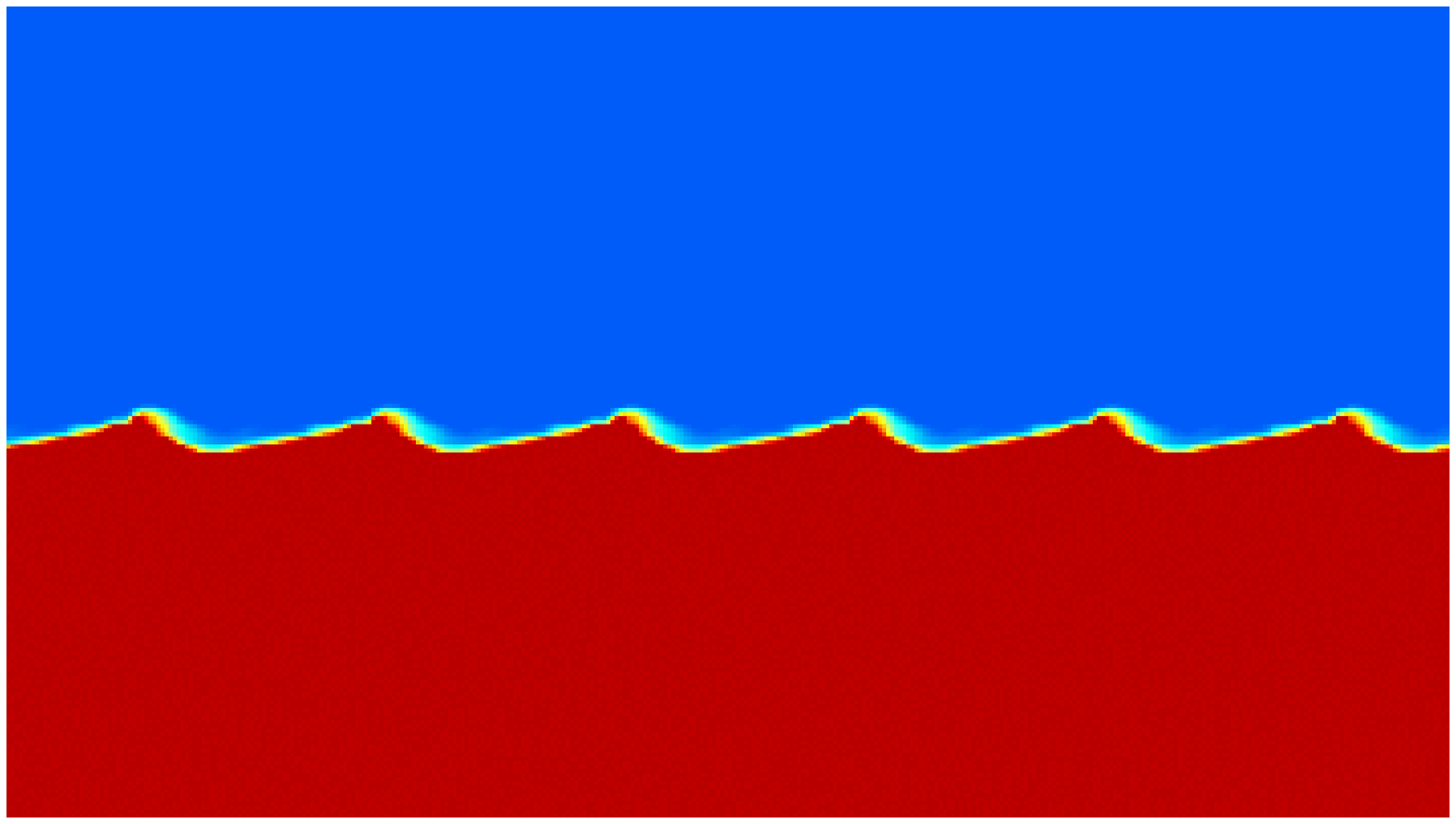,width=160pt} &
\psfig{file=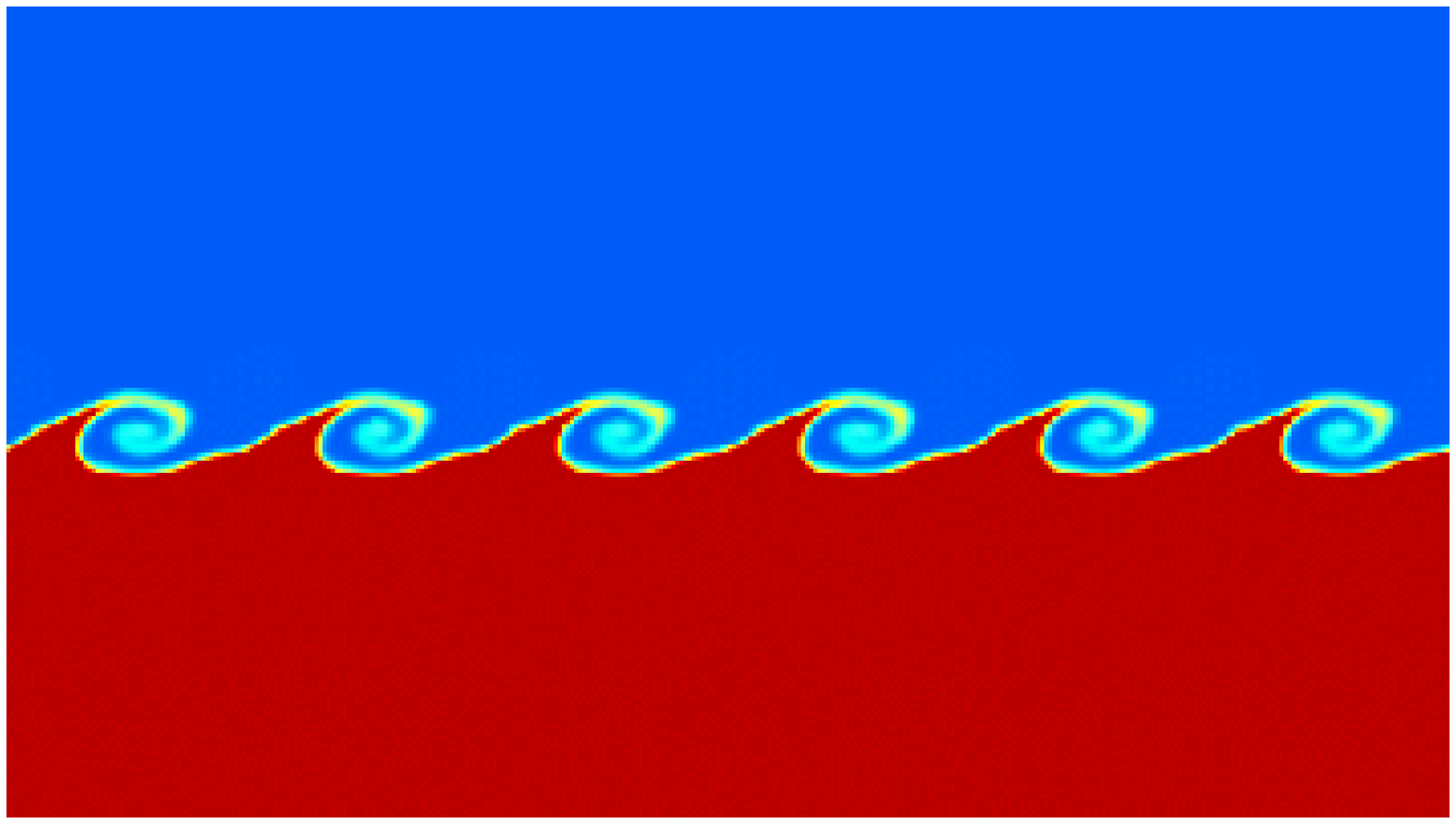,width=160pt} \\
\psfig{file=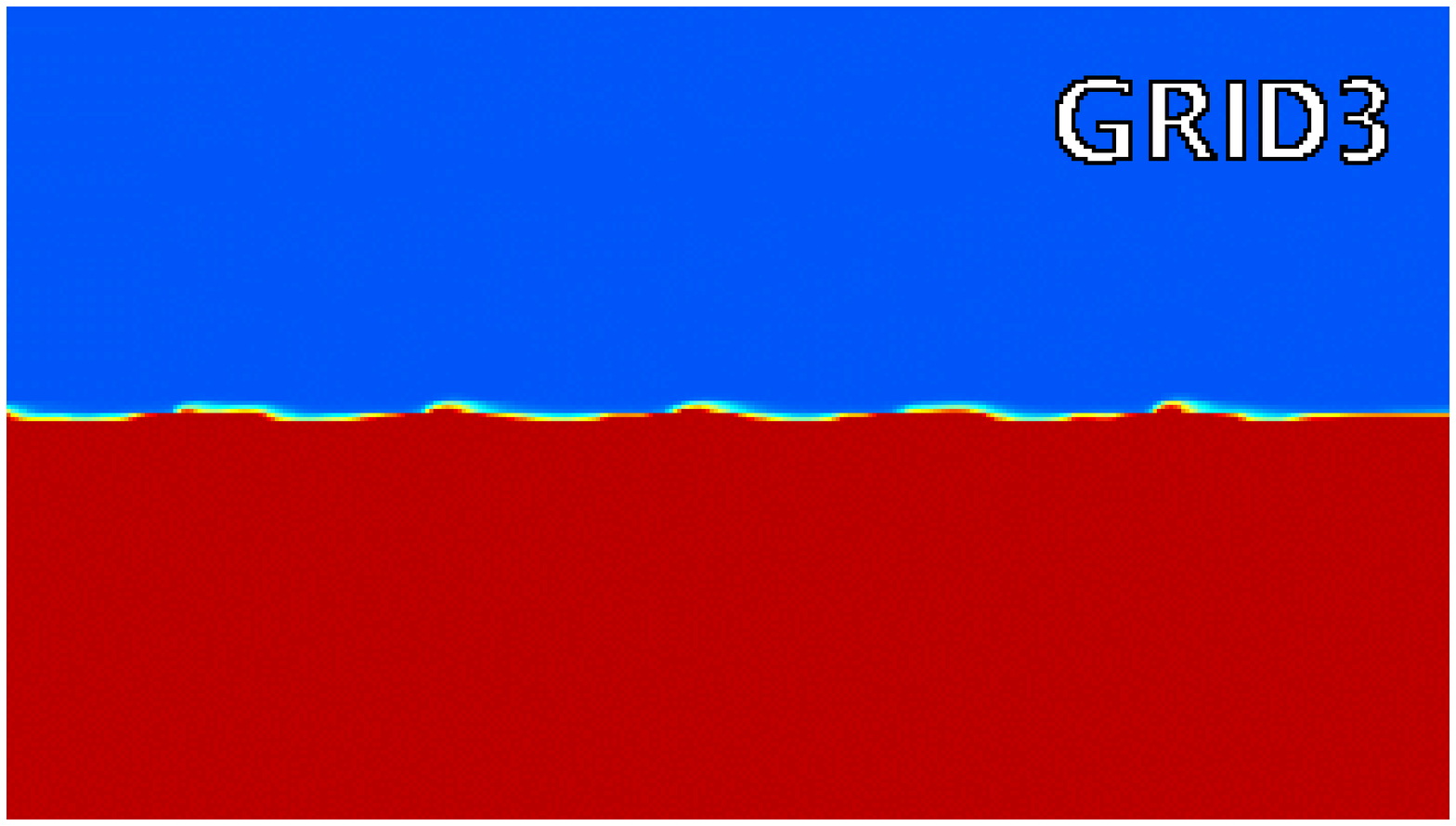,width=160pt} &
\psfig{file=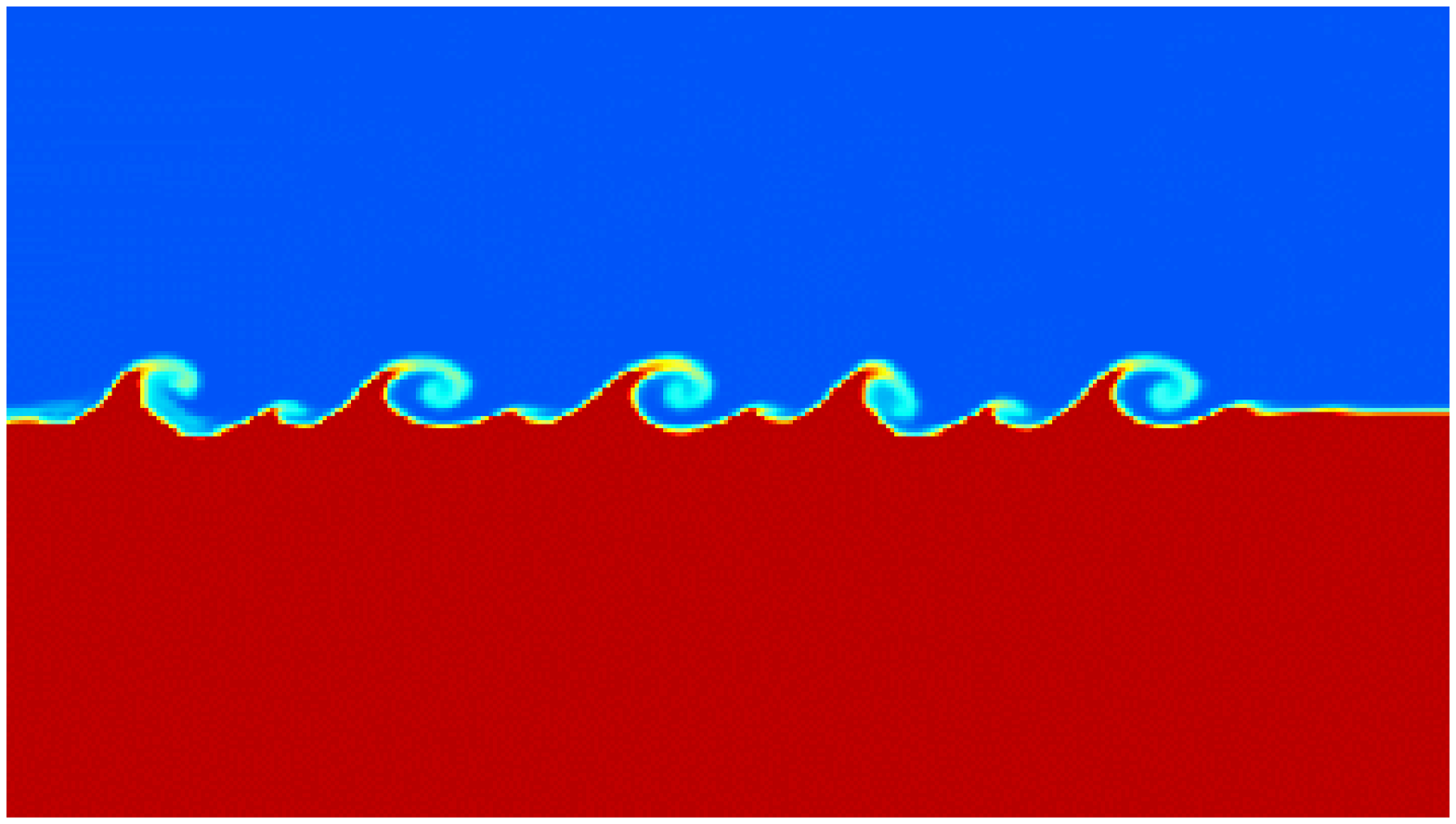,width=160pt} &
\psfig{file=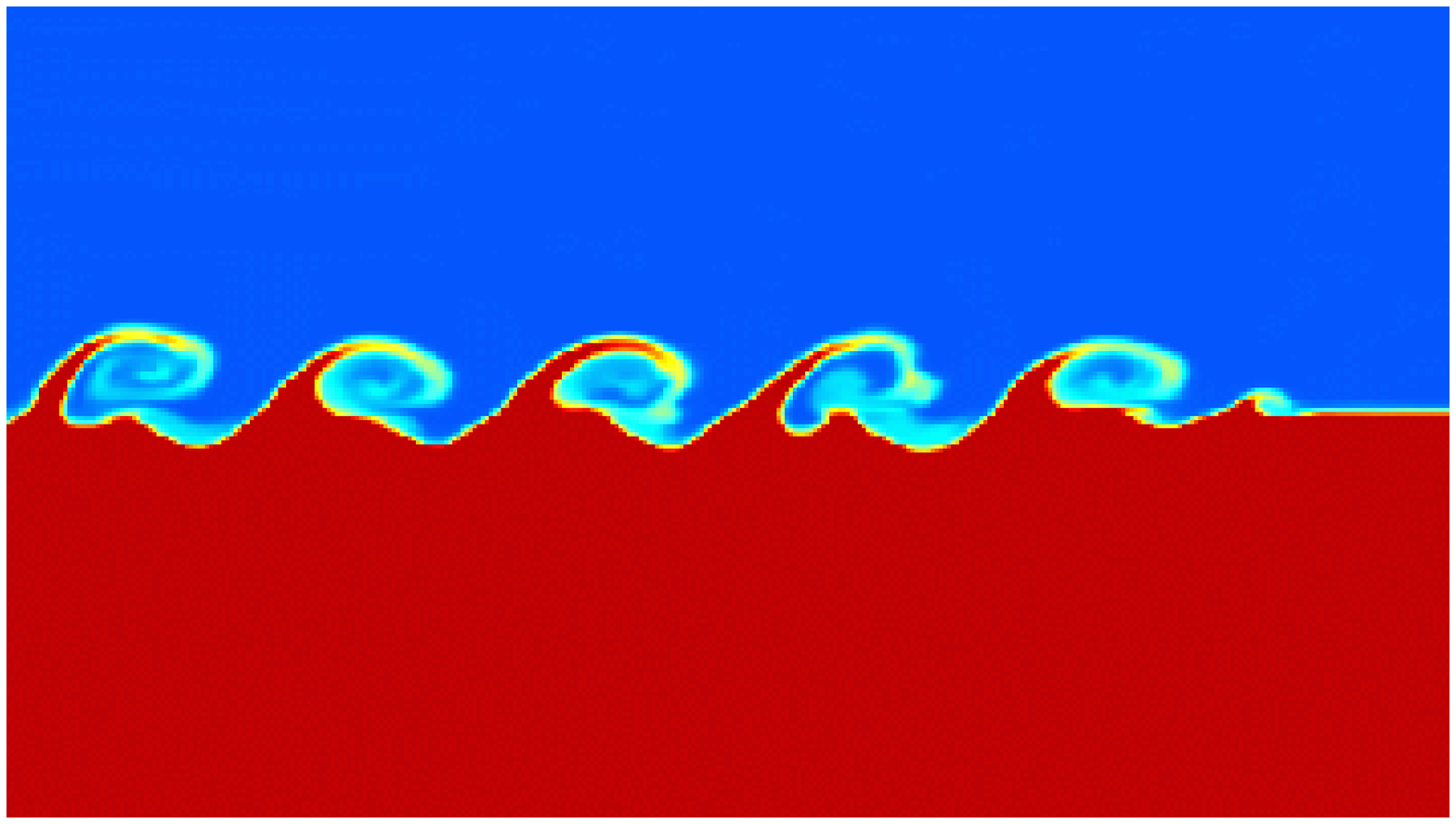,width=160pt} \\
\psfig{file=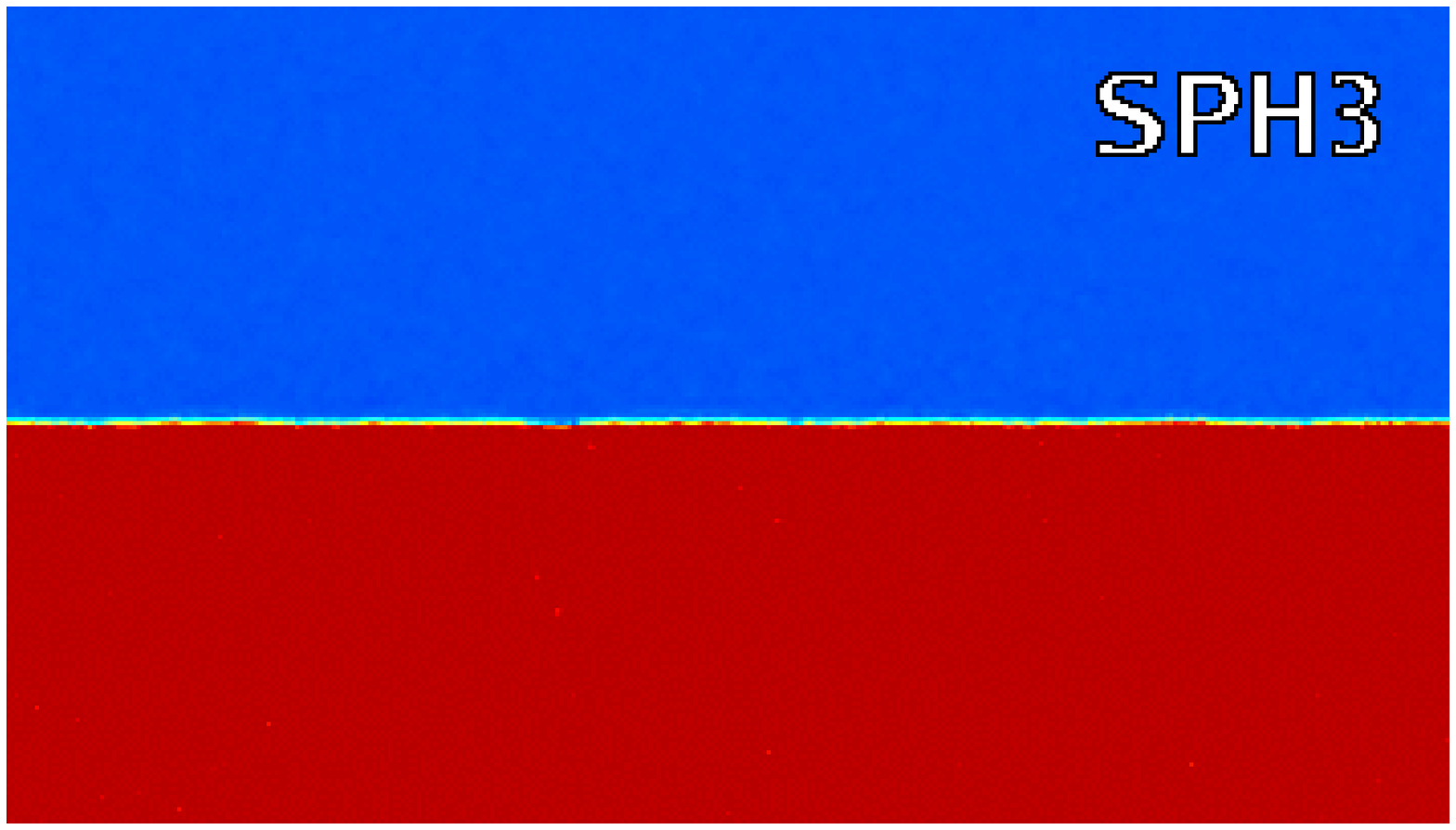,width=160pt} &
\psfig{file=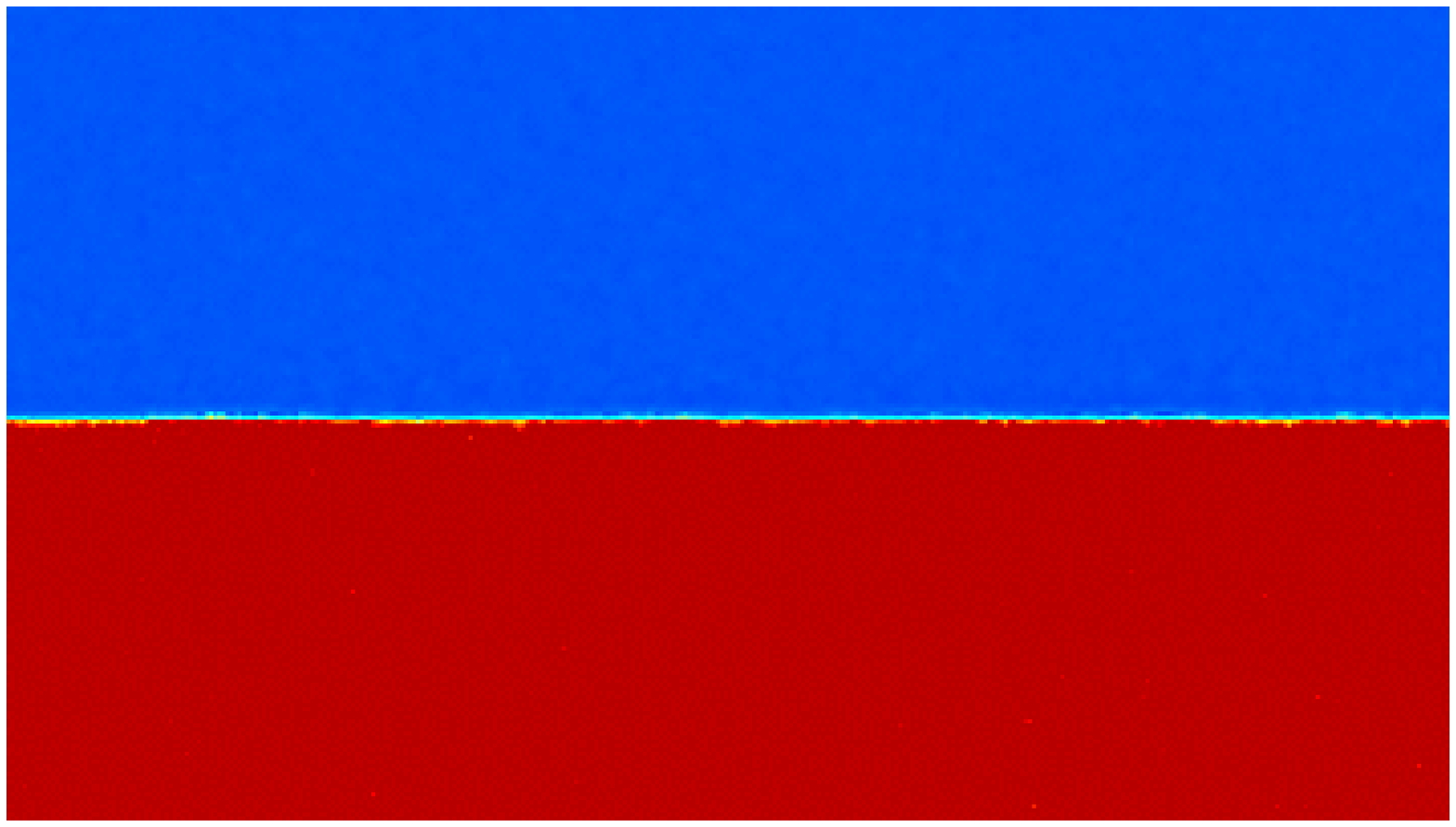,width=160pt} &
\psfig{file=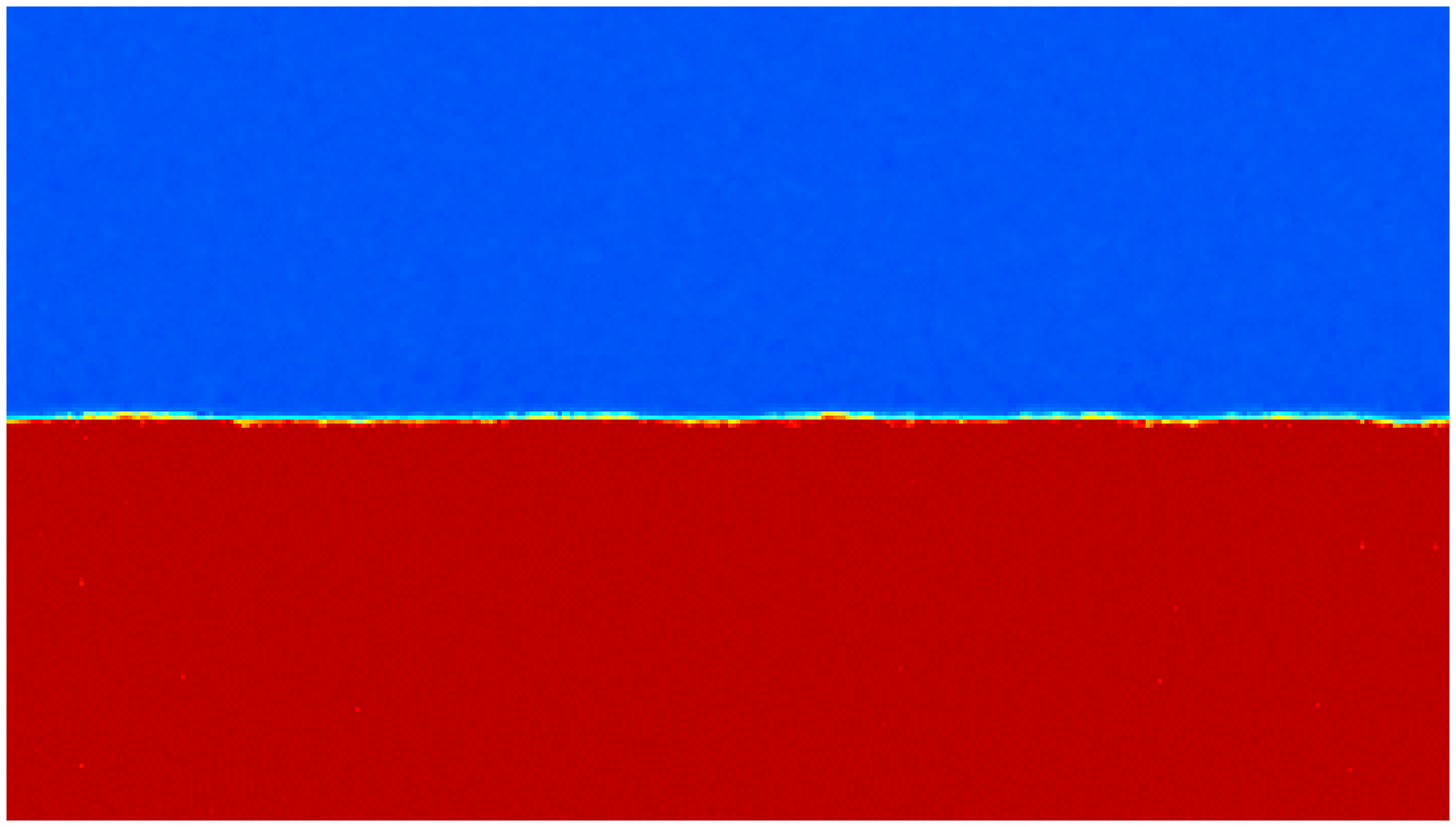,width=160pt} \\
\end{tabular}
\caption[]{Density slices of, from top to bottom, GRID1, GRID3 and SPH3. The panels show the KH simulation at $t=\tau_{\rm{KH}}/3$, $2\tau_{\rm{KH}}/3$ and $\tau_{\rm{KH}}$. It is clear from the tests that the SPH method can not resolve the KH instability. } 
\label{fig:KHI}
\end{figure*}

This set of simulations and their characteristics are summarized in Table \,\ref{table:KH} and
Fig.\,\ref{fig:KHI} shows the results, from top to bottom, GRID1, GRID3 and SPH3. We choose to show only one of the SPH results since all of these runs give the same result. GRID1 and GRID3 illustrate the difference between a highly idealised smooth setup (GRID1) and one with small scale noise (GRID3). 

GRID1 nicely produces the KH instabilities and the growth time is in excellent agreement with that expected from  Eq.\,\ref{eq:KHI}. This growth is not as clean in GRID3, which is to be expected due to local noise in density which alters the visual outcome. However the KHI is still well resolved and the growth time is comparable to the analytical expectation.

The outcome of the SPH simulation is again very different from the grids. Perturbations are damped out very quickly both in velocity and density regardless of the initial conditions, the resolution and the viscosity. We conclude that SPH in the form used in astrophysical 
simulations to-date is unable to capture dynamical instabilities such as KH when density gradients are present. As we will show in the next section, the reason for this stems from the way hydrodynamical forces are calculated in SPH in regions with strong gradients.

\subsection{Mind the gap}
\label{sect:implementation}

\begin{figure*}
\begin{tabular}{cc}
\psfig{file=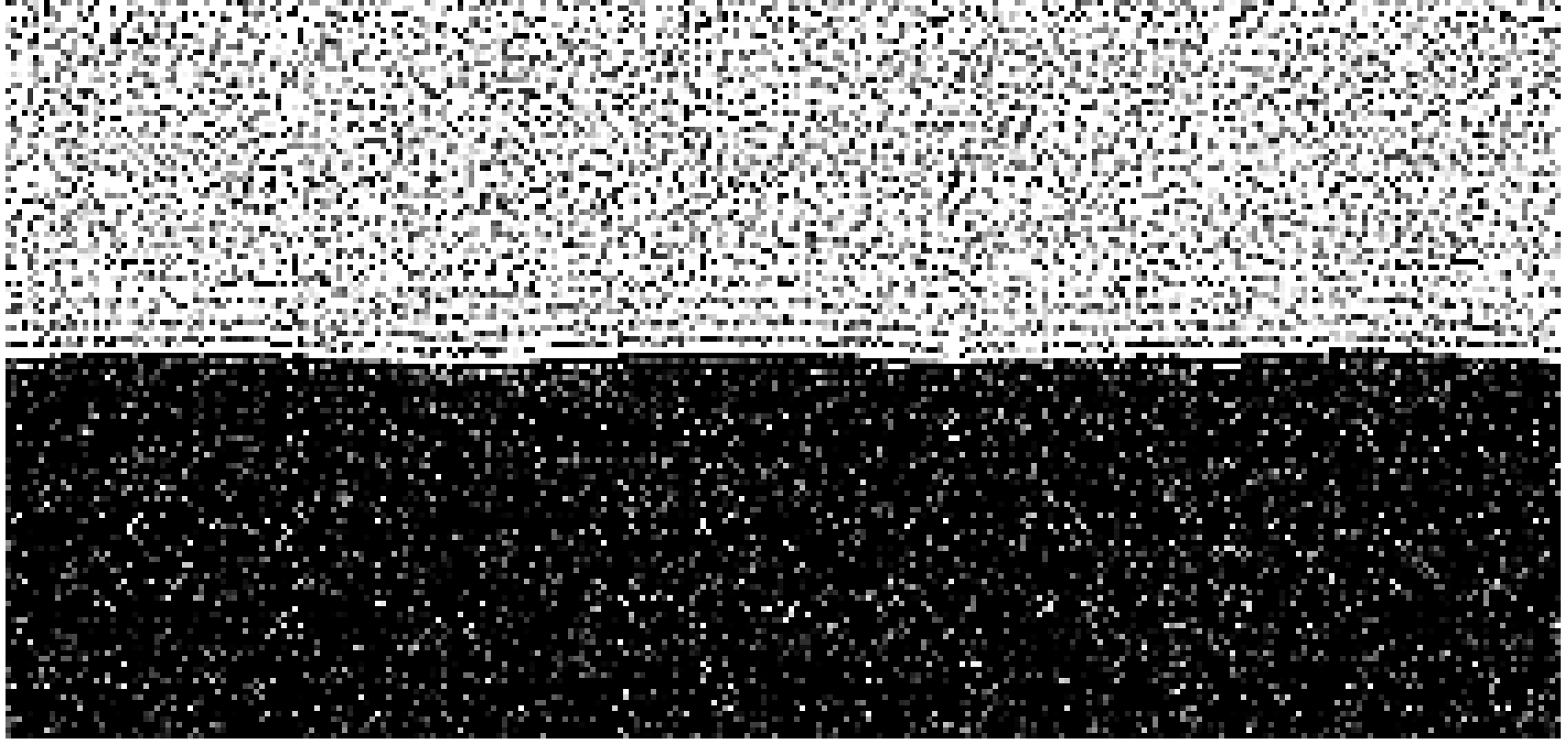,width=240pt} &
\psfig{file=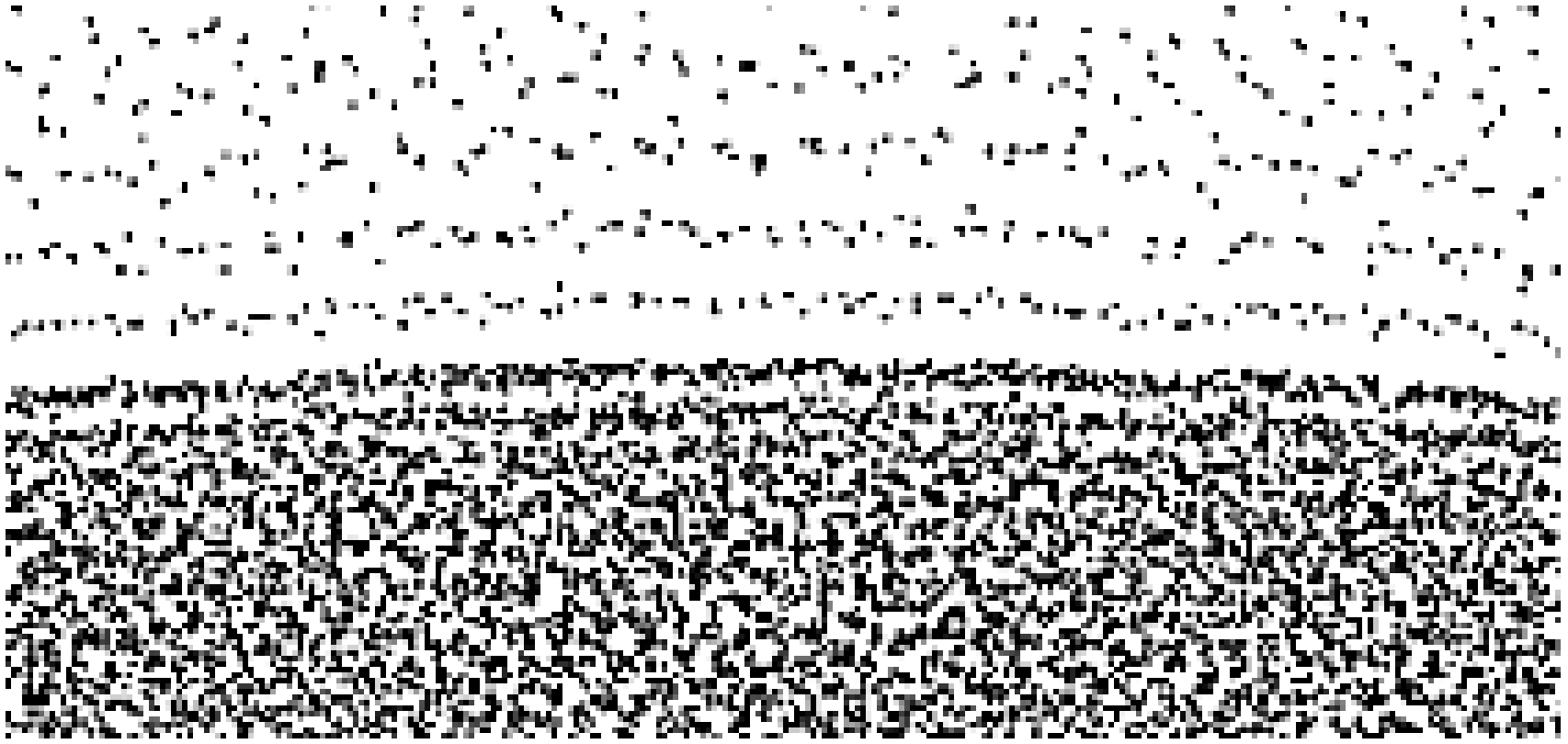,width=240pt} \\
\end{tabular}
\caption[]{A close up view of the SPH particles at the boundaries between the shearing layers (left) and closer zoom in (right) for SPH3 at $\tau_{\rm{KH}}$. We can clearly see artificial vacuum layers formed through erroneous pressure forces due to improper density calculations at density gradients. Even though the two fluids are moving relative to each other, the gap is so large that the viscosity information is not transferred between the two fluids. } 
\label{fig:vacuum}
\end{figure*}

Fig.\,\ref{fig:vacuum} shows a closeup of the SPH particles at the interface of the two fluids in SPH3 at $t=\tau_{\rm{KH}}$. There is a gap between them that has the size of the SPH smoothing kernel. This gap repeats periodically in each fluid, being smaller in the higher density fluid since the smoothing length (mean distance to the nearest 32 particles) is smaller there.
This feature is found in \emph{all} of our SPH KH simulations. It occurs very quickly and becomes more prominent with time. This phenomenon has been discussed before in the literature \citep[e.g.][]{ritchie01,tittley01}, especially in the context of numerical overcooling \citep{pearce99,thacker00,springel02,marri03} but no relation to resolving instabilities has been mentioned. 

The gap can also be clearly seen in the cloud test simulation Fig. \,\ref{fig:SPHslices}. Even though the gas is streaming with high velocity onto the leading surface of the cloud, the spurious pressure forces prevent it from making any physical contact. The reason that the cloud loses mass in the SPH simulation is due to the vacuum behind the cloud into which the cloud expands from its edges. Here the gradients become smooth and the gas can be removed by the pressure difference between the cloud and the ambient medium that streams past.

The effect can be explained in the following way:
Eq. \,\ref{eq:sphmom} is the force on each SPH particle coming from the summation over the 32 nearest neighbours. The pressure is given by $P\sim\rho T$ in the assumed case of an ideal gas. This force calculation formally assumes that temperature, and more importantly, density gradients are small within the smoothing kernel, where temperature is a quantity accumulated over time while density usually is re-estimated at each time step. When a particle from a hot low density region approaches a cold high density region it will suddenly find a lot of neighbours at the edge of the smoothing sphere within the dense medium and its density will be overestimated. This leads to, through momentum conservation, a repulsive, fictitious, force on the particle, causing it to bounce back into the low density region. This behaviour leads to the formation of a gap between the two phases of the size $\sim 2h_{ij}$, where $h_{ij}$ is the effective smoothing kernel length, either obtained by using smoothing length or smoothing kernel averaging \citep{hernquist89}, depending on the SPH implementation. Hot particles close to this gap will now have a strongly assymetric distribution of particles around them resulting in an average pressure force pointing back into the vacuum layer. Particles then travel back into the empty region and the whole process is repeated.

As mentioned above, this erroneous treatment of density contrasts has also been found to produce overcooling in galaxy formation simulations. \cite{tittley01} showed that in subsonic regimes this behaviour leads to fictitious accretion of particles on the lateral sides of gas clouds such as the simulations showed in this paper. 
Solutions to this problem has been attempted by several authors (see references above) by reformulating SPH to more accurately treat the particle interactions at steep boundaries. While this seems to remove the vacuum layer to some extent, it is unclear how this will affect the simulations discussed here. Possible solutions to the gap problem, such as modifying the viscosity weighting kernel, will be presented in a follow up paper by \cite{agertz07}.

\begin{figure*}
\begin{tabular}{cc}
\psfig{file=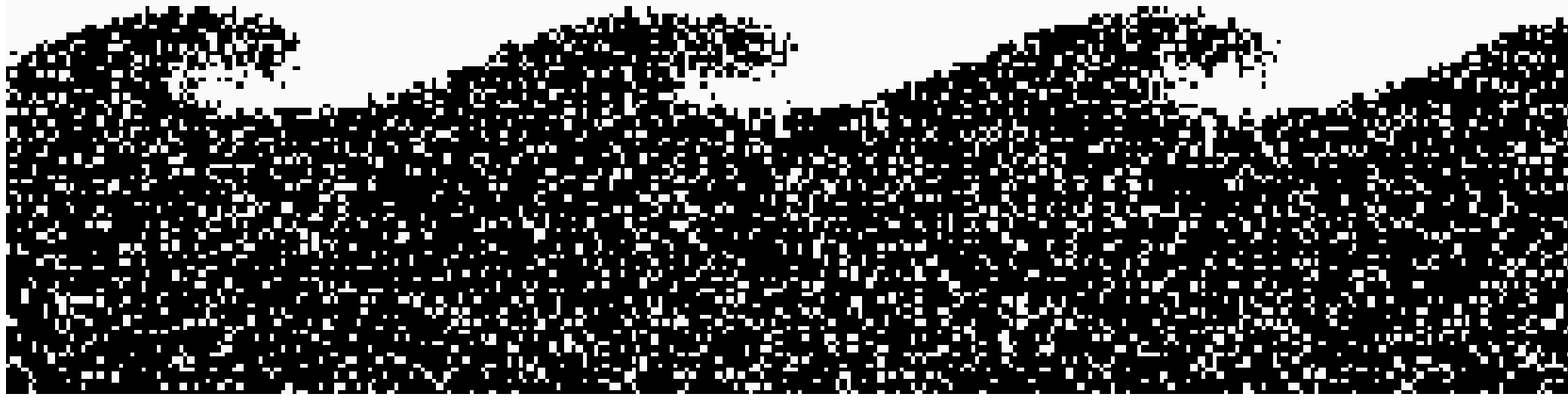,width=240pt} &
\psfig{file=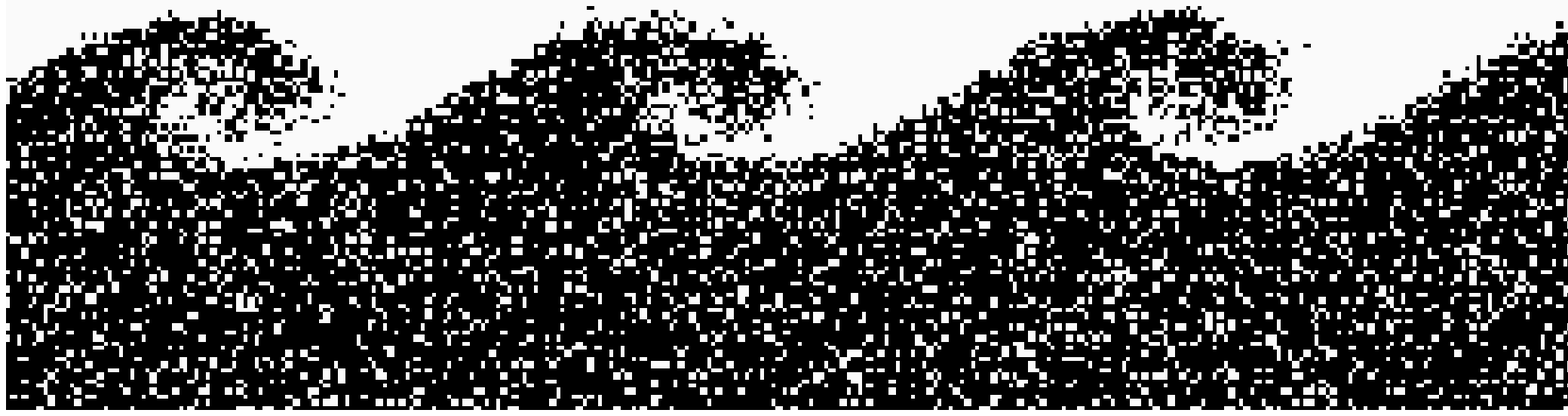,width=240pt} \\
\end{tabular}
\caption[]{A zoom in of the SPH particles at the boundaries between the shearing layers for the isodensity SPH run with standard viscosity (left) and low viscosity (right) at $\tau_{\rm{KH}}$. The black and white region are particles that belonged to the initially separated shearing layers. We clearly see that the growth of the KHI in the standard implementation of SPH, and even stronger for the low viscosity version.  } 
\label{fig:KHSPH}
\end{figure*}

That the gap is the origin of instability supression becomes even more aparent by studying the KHI using a density contrast $\chi=1$, in which the gap can not form. With this vanishing density gradient, SPH is able to capture the KHI, see Fig. \ref{fig:KHSPH}. The left panel shows the KHI at $t=\tau_{\rm KH}$ for the standard $\alpha=1.0$, $\beta=2.0$ setting and the right panel shows the same timestep but using $\alpha=0.01$ and $\beta=1.0$. The less evolved standard viscosity simulation points out the effects of viscosity discussed in section \ref{sect:AV}. Similar results have been recently found by \cite{junk06}.

\section{Summary}
\label{sec:summary}

In this paper we have carried out hydrodynamical simulations of
a cold gas cloud interacting with an ambient hot moving gas using state of the art simulations codes. Striking differences were found between the two main techniques for simulating fluids.
While grid codes are able to resolve and treat dynamical instabilities and mixing, these processes are poorly or not at all resolved by the current SPH techniques. We show that the reason for this is that SPH, at least in the standard usage and formulation, inaccurately handles situations where density gradients are present. In these situations, SPH particles of low density close to high density regions suffer erroneous pressure forces due to the assymetric density within the smoothing kernel. This causes a gap between regions of high density contrast. 

This behaviour has implications for many astrophysical situations. For example the stripping of gas from galaxies moving through a gaseous medium has already been discussed in the literature. The origin of disk galaxies is an important unsolved problem. Perhaps the inability to disrupt accreting gas clouds is one reason why numerical calculations have failed to produce pure disk systems. Simulating star formation regions and feedback processes also relies on the correct ability to model turbulence and interacting multiphase fluids.

It should be noted that the behaviour of the grid and SPH mehods agree on timescales shorter than those of typical dynamical instabilities such as the Kelvin-Helmholtz and Rayleigh-Taylor instabilities. In our specific test of a cold cloud engulfed in a hot wind, there is good agreement in the early gas stripping phase occuring due to pressure differences arising in the Bernoulli zones. As soon as the large scale instabilities have grown, the results of the different methods diverge. There are several possible solutions to this behaviour in SPH calculations which we will explore in a seperate work.

\section*{Acknowledgements}
We acknowledge support from the European Science Foundation who funded an
exploratory workshop in Wengen 2004 at which these tests were first discussed. 
FM and LM acknowledge support by the Swiss Institute of Technology through
a Zwicky Prize Fellowship. 
OA would like to thank Alessandro Romeo and Peter Englmaier for valuable discussions.
AJG acknowledge the support from the Polish Ministry of Science through
the grant 1P03D02626 and from the European Community's Human Potential
Programme through the contract HPRN-CT-2002-00308, PLANETS. The AMR software
(FLASH) used in this work was in part developed by the DOE-supported
ASC / Alliance Centre for Astrophysical Thermonuclear Flashes at the
University of Chicago. The FLASH calculations were performed at the
Interdisciplinary Centre for Mathematical and Computational Modeling in
Warsaw, Poland.

\bibliographystyle{/benutzer/theorie/agertz/LATEX/mn2e}
\bibliography{blob.bbl}

\end{document}